\documentclass[iop]{emulateapj}

\def\lapp{\ifmmode\stackrel{<}{_{\sim}}\else$\stackrel{<}{_{\sim}}$\fi}
\def\gapp{\ifmmode\stackrel{>}{_{\sim}}\else$\stackrel{>}{_{\sim}}$\fi}
\usepackage{multirow}
\usepackage{color}
\usepackage{amsmath}
\usepackage{soul}

\newcommand{\alphamag}{\alpha_{\rm mag}}
\newcommand{\betaobs}{\beta_{\rm obs}}
\newcommand{\thetaj}{\theta_{j}}
\newcommand{\phaseref}{\phi_{\rm 0}}

\newcommand{\lj}{L_j}

\begin{document}

\title{Deep NuSTAR and Swift Monitoring Observations of the Magnetar 1E~1841$-$045}

\author{
Hongjun An\altaffilmark{1,2},
Robert~F.~Archibald\altaffilmark{1},
Romain~Hasco{\"e}t\altaffilmark{3},
Victoria~M.~Kaspi\altaffilmark{1},
Andrei~M.~Beloborodov\altaffilmark{3},
Anne~M.~Archibald\altaffilmark{4},
Andy~Beardmore,\altaffilmark{5},
Steven~E.~Boggs\altaffilmark{6},
Finn~E.~Christensen\altaffilmark{7},
William~W.~Craig\altaffilmark{6,8}, 
Niel~Gehrels\altaffilmark{9},
Charles~J.~Hailey\altaffilmark{3},
Fiona~A.~Harrison\altaffilmark{10},
Jamie~Kennea\altaffilmark{11},
Chryssa~Kouveliotou\altaffilmark{12},
Daniel~Stern\altaffilmark{13},
George~Younes\altaffilmark{12},
and William~W.~Zhang\altaffilmark{14}\\
}
\affil{
{\small $^1$Department of Physics, McGill University, Montreal, QC H3A 2T8, Canada}\\
{\small $^2$Department of Physics/KIPAC, Stanford University, Stanford, CA 94305-4060, USA}\\
{\small $^3$Columbia Astrophysics Laboratory, Columbia University, New York, NY 10027, USA}\\
{\small $^4$ASTRON, The Netherlands Institute for Radio Astronomy, Postbus 2, 7990 AA, Dwingeloo, The Netherlands}\\
{\small $^5$Department of Physics and Astronomy, University of Leicester, University Road, Leicester LE17RH, UK}\\
{\small $^6$Space Sciences Laboratory, University of California, Berkeley, CA 94720, USA}\\
{\small $^7$DTU Space, National Space Institute, Technical University of Denmark, Elektrovej 327, DK-2800 Lyngby, Denmark}\\
{\small $^8$Lawrence Livermore National Laboratory, Livermore, CA 94550, USA}\\
{\small $^9$Astrophysics Science Division, NASA Goddard Space Flight Center, Greenbelt, MD 20771 USA}\\
{\small $^{10}$Cahill Center for Astronomy and Astrophysics, California Institute of Technology, Pasadena, CA 91125, USA}\\
{\small $^{11}$Department of Astronomy and Astrophysics, 525 Lab, Pennsylvania State University, University Park, PA 16802, USA}\\
{\small $^{12}$Space Science Office, ZP12, NASA Marshall Space Flight Center, Huntsville, AL 35812, USA}\\
{\small $^{13}$Jet Propulsion Laboratory, California Institute of Technology, Pasadena, CA 91109, USA}\\
{\small $^{14}$Goddard Space Flight Center, Greenbelt, MD 20771, USA}\\
}
\begin{abstract}
We report on a 350-ks {\it NuSTAR} observation of the magnetar 1E~1841$-$045
taken in 2013 September. During the observation,
{\it NuSTAR} detected six bursts of short duration, with $T_{\rm 90}\lapp 1\rm \ s$.
An elevated level of emission tail is detected
after the brightest burst, persisting for $\sim$1\,ks.
The emission showed a power-law decay with a temporal index of 0.5 before returning
to the persistent emission level. The long observation also
provided detailed phase-resolved spectra of the persistent
X-ray emission of the source. By comparing the persistent spectrum with
that previously reported, we find that the source hard-band emission
has been stable over approximately 10 years.
The persistent hard X-ray emission
is well fitted by a coronal outflow model, where $e^{\pm}$ pairs
in the magnetosphere upscatter thermal X-rays. Our fit of
phase-resolved spectra allowed us to estimate the angle
between the rotational and magnetic dipole axes of the magnetar,
$\alphamag = 0.25$, the twisted magnetic flux,
$2.5 \times 10^{26}\rm \ G\ cm^2$, and the power released
in the twisted magnetosphere, $\lj = 6\times10^{36}\rm \ erg\ s^{-1}$.
Assuming this model for the hard X-ray spectrum,
the soft X-ray component is well fit by a two-blackbody model,
with the hotter blackbody consistent with the footprint of
the twisted magnetic field lines on the star.
We also report on the 3-year {\it Swift} monitoring observations obtained
since 2011 July. The soft X-ray spectrum remained stable during this period,
and the timing behavior was noisy, with large timing residuals.
\end{abstract}


\section{Introduction}
Magnetars are neutron stars to have emission that is powered
by the decay of their intense magnetic fields \citep[][]{dt92,td96}.
The magnetic field strengths inferred from the spin-down parameters are
typically greater than $10^{14}\rm \ G$ \citep[e.g.,][]{vg97,kds+98},
though there are several sources with lower inferred field strengths
\citep[e.g., SGR~0418$+$5729, Swift~J1822.3$-$1606;][]{ret+10, skc14}.
There are 28 magnetars discovered to date, including six candidates
\citep[see][]{ok14}.\footnote{See the online magnetar catalog for
an up-to-date compilation of known magnetar properties,
http://www.physics.mcgill.ca/$\sim$pulsar/magnetar/main.html}

Short X-ray bursts are often detected from magnetars.
The bursts have a variety of morphologies, spectra, and energies,
and are thought to be produced
 by crustal or magnetospheric activity \citep[][]{tlk02}.
Interestingly, some bursts are followed by a long emission tail
while others are not. It has been suggested that the energy
in the burst and the integrated energy in the burst tail
are correlated \citep[for SGR~1900$+$14 and SGR~1806$-$20,][]{lwg+03,gwk+11},
which may imply that their relative strengths show a
narrow distribution. \citet{wkg+05} suggested a bimodal distribution
for the relative strengths across magnetars, implying two distinct physical
mechanisms for the bursts. In addition to short X-ray bursts,
giant flares and dramatic increases in the persistent emission
over days to months have been also observed in some sources
\citep[see][for reviews]{wt06,k07,re11,m13}.

The persistent emission of magnetars in the X-ray band below $\sim$10~keV
is dominated by thermal emission and is often modeled with two blackbodies
from two hot regions on the surface, or with a surface blackbody plus
a power law resulting from magnetospheric reprocessing
\citep[][]{tlk02,zrt+09}. Some magnetars also show significant emission
in the hard X-ray band above $\sim$10~keV \citep[][]{khh+06} which
is believed to be produced in the magnetosphere
\citep[][]{tb05, hh05, bh07, bt07}.
Recently, \citet{b13} proposed a coronal outflow model for the hard X-ray
emission. The model makes specific predictions for phase-resolved
spectra which can be tested by observations. It has been applied to 
four magnetars with available phase-resolved data above 10~keV, and in all 
cases the model was found consistent with the data
\citep[][]{ahk+13, hbd14, vhk+14}.

The magnetar 1E~1841$-$045 has a surface dipolar magnetic-field strength of
$B \equiv 3.2 \times 10^{19}(P\dot P)^{1/2}\ \rm \ G=6.9 \times 10^{14}\rm \ G$,
estimated from the spin period and the spin-down rate of $P=11.8$~s and
$\dot P=4\times10^{-11}\rm \ s \ s^{-1}$, respectively,
assuming the standard vacuum dipole spin-down model.
The source has previously shown occasional X-ray bursts with energies
of $\sim10^{38}\rm \ erg$ assuming a distance of 8.5~kpc
\citep{ks10, lkg+11,dk14}. No tails or significant enhancement
in the persistent emission were observed following the bursts.
\citet{ks10} reported detection of emission lines $\gapp$20\,keV in the burst spectrum
measured with the {\it Swift} Burst Alert Telescope (BAT).
Furthermore, the authors argued that the source brightened,
and the emission became softer after the burst activity.
However, \citet{lkg+11} did not find any emission line,
flux enhancement or spectral changes at the burst epoch,
in contradiction with the results of \citet{ks10}.

1E~1841$-$045 is one of the brightest magnetars in the hard X-ray band
above 10~keV. Its persistent emission has been studied by \citet{khh+06}
and \citet{ahk+13}. The measured photon index in the hard X-ray band
is $\Gamma\sim 1.3$, and the pulsed fraction was reported to increase
with photon energy. \citet{ahk+13} applied the coronal outflow model of
\citet{b13} to a 50-ks {\it NuSTAR} observation and constrained the hard X-ray
emission geometry to two possible solutions. Furthermore, an interesting
feature was found in the pulse profile in the 24--35~keV band,
which may be associated with a spectral feature in this band.

In this paper, we further investigate the properties of persistent emission
of 1E~1841$-$045 using a new 350-ks {\it NuSTAR} observation and
3-year monitoring observations by {\it Swift}. Fortuitously, the source
was actively bursting during the {\it NuSTAR} observation,
which provided an opportunity to study the bursts in addition to
the persistent emission. We describe the {\it NuSTAR}
and archival {\it Chandra}, {\it XMM-Newtor} and {\it Swift} observations
used in this paper in Section~\ref{sec:obs} and present the results of
the {\it NuSTAR} data analysis in Section~\ref{sec:ana}.
We then present the {\it Swift} monitoring observations and the data analysis
in Sections~\ref{sec:swiftobs} and \ref{sec:swiftmonitoring}.
We discuss the results of data analyses in Section~\ref{sec:disc}
and summarize our main conclusions in Section~\ref{sec:concl}.
\medskip

\newcommand{\markzz}{\tablenotemark{a}}
\newcommand{\markza}{\tablenotemark{b}}
\newcommand{\markzb}{\tablenotemark{c}}
\begin{table}[t]
\vspace{-0.0in}
\begin{center}
\caption{Summary of observations used in this work
\label{ta:obs}}
\vspace{-0.05in}
\scriptsize{
\begin{tabular}{cccccccc} \hline\hline
Observatory	&Obs. ID	& Obs. Date	& Exposure	& Mode\markzz	\\ 
		&		& (MJD)		& (ks)		&	\\ \hline 
{\em Chandra}	& 730		& 51754.3	& 10.5		& CC \\ 
{\em XMM-Newton}& 0013340101	& 52552.2	& 3.9		& FW/LW   \\ 
{\em XMM-Newton}& 0013340102	& 52554.2	& 4.4		& FW/LW   \\ 
{\em Chandra}	& 6732		& 53946.4	& 24.9		& TE   \\ 
{\em Swift}\markza	& 00080220003	& 56241.3	& 17.9		& PC  \\ 
{\em Swift}\markza	& 00031863050	& 56551.3	& 4.3		& WT  \\ 
{\em Swift}\markza	& 00080220004	& 56556.5	& 1.9		& PC  \\ 
{\em NuSTAR}	& 30001025002	& 56240.9	& 48.6		& $\cdots$  \\  
{\em NuSTAR}	& 30001025004	& 56540.3	& 35.9		& $\cdots$  \\  
{\em NuSTAR}	& 30001025006	& 56542.4	& 77.9		& $\cdots$  \\  
{\em NuSTAR}	& 30001025008	& 56547.1	& 85.7		& $\cdots$  \\  
{\em NuSTAR}	& 30001025010	& 56549.6	& 53.3		& $\cdots$  \\  
{\em NuSTAR}	& 30001025012	& 56556.5	& 100.7		& $\cdots$  \\ \hline 
\end{tabular}}
\end{center}
\vspace{0.0mm}
$^{\rm a}${CC: Continuous Clocking, FW: Full Window, LW: Large Window, PC: Photon Counting,
TE: Timed Exposure, WT: Windowed Timing. MOS1,2/PN for {\em XMM-Newton}.}\\
$^{\rm b}${{\it Swift} observations used for spectral analysis in Section~\ref{sec:phavspec}.
Results of analysis on a larger {\it Swift} monitoring dataset are presented in Section~\ref{sec:swiftobs}.}\\
\vspace{-5.0 mm}
\end{table}

\section{Observations}
\label{sec:obs}
1E~1841$-$045 was observed with {\it NuSTAR} \citep{hcc+13} between 2013 September 5
and 21 in a series of exposures with durations 40--100 ks with
a total net exposure of 350 ks. Two X-ray bursts were detected with
the {\it Fermi} Gamma-Ray Burst Monitor (GBM) \citep[][]{cxc14, paf+14}
on 2013 September 13, during the {\it NuSTAR} observation period.
Fortunately, the bursts were also recorded in the {\it NuSTAR} data.
In addition to the {\it Fermi} reported bursts, {\it NuSTAR} serendipitously
detected several more bursts. We report on the bursts in Section~\ref{sec:burstana}.
To study the persistent emission of the source, we also analyzed archival
observations made previously with {\it NuSTAR}, {\it Chandra}, {\it XMM-Newton}
and {\it Swift} to have better statistics and to constrain
the persistent properties of the source in the soft band.
The observations used in this study are listed in Table~\ref{ta:obs}.
Note that all the soft-band observations and the first {\it NuSTAR} observation
(Obs. ID 30001025002) were reported previously \citep[][A13 hereafter]{ahk+13}.

The {\it NuSTAR} data were processed with {\tt nupipeline} 1.3.1 along with
CALDB version 20131223. We used default filters except for {\tt PSDCAL}
for which we used {\tt PSDCAL=NO}. The {\tt PSDCAL} filter is
for laser metrology calibration of the relative positions of
the optics and detectors.
This observation was affected by times when the metrology laser
went out of range.
By not using the default, the pointing accuracy
may be slightly degraded, but good time intervals (GTI), and hence
exposure time, increase.\footnote{See http://heasarc.gsfc.nasa.gov/docs/nustar/analysis/nustar\\\_swguide.pdf for more details}
We note this situation is unusual and specific to this observation.
We verified that the analysis results described below
do not change depending
on the {\tt PSDCAL} filter setting. However, we note that some of
the burst data could be retrieved only with {\tt PSDCAL=NO}.

We also analyzed archival {\it XMM-Newton} and {\it Chandra}
observations (see Table~\ref{ta:obs}). The {\it XMM-Newton} data were
processed with Science Analysis System (SAS) 12.0.1, and the {\it Chandra} data
were reprocessed using chandra\_repro of CIAO 4.4 along with CALDB 4.5.3.
We further processed the cleaned data for analysis as described below.
Uncertainties below are at the 1$\sigma$ confidence level unless stated otherwise.

\medskip
\section{Data Analysis and Results for the {\it NuSTAR} observations}
\label{sec:ana}
In this Section, we present data analysis results for bursts and persistent
emission measured with the observations in Table~\ref{ta:obs}
(Sections~\ref{sec:burstana}-\ref{sec:phresspec}).
We fit the persistent and phase-resolved spectra with
the coronal outflow model and show the results in Section~\ref{sec:bmodel}.

\subsection{Burst Analysis}
\label{sec:burstana}
\subsubsection{Temporal and Spectral Properties of the Bursts}
\label{sec:bursttiming}

\newcommand{\markbz}{\tablenotemark{a}}
\newcommand{\markcz}{\tablenotemark{b}}
\newcommand{\markdz}{\tablenotemark{c}}
\begin{table*}[ht]
\vspace{0.0 mm}
\begin{center}
\caption{Deadtime-corrected Properties of {\em NuSTAR}-detected
Bursts from 1E~1841$-$045
\label{ta:bursts}}
\scriptsize{
\begin{tabular}{cccccccccccccccc} \hline\hline
Burst & $T_{\rm 0}$ & $\phi$\markbz & $T_{\rm r}$ & $T_{\rm f}$ & $A$ & $C_{\rm 1}$ & $C_{\rm 2}$ & $T_{\rm 90}$\markcz & $N_{\rm evt}$ & $\Delta_{\rm post-pre}$\\
        & (day)&  & (s)& (s)    & (cps) & (cps) & (cps) & (s) & (cts) & (cts) \\ \hline
1 & 0.35836545 & $0.2266^{+0.0003}_{-0.0002}$ & $0.007^{+0.001}_{-0.001}$ & $0.047^{+0.017}_{-0.019}$ & $1740^{+760}_{-600}$ & $12^{+2}_{-2}$ & $<16$ & $0.12^{+0.04}_{-0.05}$ & 5 & $\cdots$ \\ 
2 & 0.35837490 & $0.2958^{+0.0002}_{-0.0002}$ & $0.006^{+0.002}_{-0.002}$ & $0.09^{+0.01}_{-0.01}$ & $700^{+100}_{-90}$ & $5.5^{+1.6}_{-1.4}$ & $5.9^{+1.8}_{-1.6}$ & $0.22^{+0.03}_{-0.03}$ & 61 & $75\pm41$ \\
3 & 0.60981692 & $0.5583^{+0.0002}_{-0.0001}$ & $0.014^{+0.002}_{-0.001}$ & $0.011^{+0.003}_{-0.002}$ & $2010^{+460}_{-410}$ & $12^{+2}_{-2}$ & $5^{+3}_{-3}$ & $0.057^{+0.011}_{-0.007}$ & 27 & $16\pm40$ \\ 
4 & 7.27821493 & $0.4463^{+0.0002}_{-0.0002}$ & $0.0053^{+0.0016}_{-0.0014}$ & $0.052^{+0.02}_{-0.02}$ & $840^{+300}_{-240}$ & $16^{+2}_{-2}$ & $<3.4 $ & $0.13^{+0.05}_{-0.05}$ & 11 & $26\pm40$ \\ 
5 & 8.62801288 & $0.1075^{+0.0001}_{-0.0001}$ & $0.0090^{+0.0003}_{-0.0003}$ & $0.0184^{+0.0004}_{-0.0004}$ & $67000^{+70000}_{-60000}$ & $15^{+1}_{-1}$ & $<3$ & $0.0631^{+0.0007}_{-0.0007}$ & 22 & $211\pm43$ \\ 
6 & 8.759684723 & $0.83672^{+0.00001}_{-0.00001}$ & $<0.0006$ & $0.0249^{+0.0016}_{-0.0018}$ & $8000^{+1200}_{-1100}$ & $14^{+2}_{-1}$ & $<4$ & $<0.059$ & 31 & $7\pm41$ \\ \hline 
\end{tabular}}
\end{center}
\vspace{-1.0 mm}
\footnotesize{
{\bf Notes.}
Parameters for the short timescale light curve.
$T_{\rm 0}$ is the burst arrival time and is days since MJD~56540
(barycentric dynamical time).
$T_{\rm r,f}$ are the rising and falling times for the burst light curves,
$A$ is the peak count rate, $C_{\rm 1,2}$ are the constant level
of the light curves before and after the bursts (see Equation~\ref{eqn:lightcurve}),
$T_{\rm 90}$ is the time interval which includes 90\% of the burst counts
estimated with the exponential functions,
$N_{\rm evt}$ is the number of events within $T_{\rm 90}$,
and $\Delta_{\rm post-pre}$ is the difference in numbers of photons contained
in the pre- and the post-burst 200-s intervals.}\\
$^{\rm a}${Spin phase corresponding to $T_{\rm 0}$,
where phase zero is defined at the pulse minimum ($T_{\rm ref}=56540.32899020$~MJD),
the same as that for the timing analysis in Section~\ref{sec:PPPF}.}\\
$^{\rm b}${ Since $T_{\rm 90}$ for the
whole burst is not always well defined because the constants $C_{\rm 1,2}$ are different
before and after the burst peak, $T_{\rm 90}$'s for
the rising and the falling function were calculated separately and
then summed to obtain that for the burst. When only an upper limit
is available for $T_{\rm r}$ or $T_{\rm f}$, we
used the upper limit to calculate $T_{\rm 90}$ and show it without uncertainties.}\\
\vspace{-3.0 mm}
\end{table*}

We performed a comprehensive search for bursts in the {\it NuSTAR} light curves.
We extracted event time series, applied the barycenter
correction using the position
R.A.=$18^{\rm h}41^{\rm m}19^{\rm s}.343$,
Decl.=$-04^{\circ}56'11\farcs$8 \citep[J2000;][]{d05},
binned the light curves with a variety of bin sizes
ranging from 0.01 to 10 s, and searched for time bins which contained
more counts than expected above background including source
persistent emission using Poisson statistics.
The background was extracted in an interval 10-pulse periods long
using the same extraction region,
just prior to the time bin being considered. In total,
we found 7 time bins which are significantly above the mean level
after considering the number of trials.
Note that two of the seven significant bins turned out to be produced by
one burst (burst 5; see below), hence we found six bursts during our observations.
The significance of the bursts is high ($p<10^{-10}$)
but only on short timescales, e.g., $\lapp$0.1~s.
We list the burst times in Table~\ref{ta:bursts}.
Note that bursts 5 and 6 were previously reported based on {\it Fermi}
GBM detections \citep[][]{cxc14, paf+14}.

We note that some of the bursts may not be fully sampled due to high count rates.
Specifically, the maximum count rate of {\it NuSTAR} detectors is $\sim$400~cps
limited by a deadtime of 2.5~ms per event. In addition,
an event that was detected in a very short elapsed time from the previous event
is regarded as background and filtered out during the standard pipeline process.
We investigated these effects by looking into the elapsed time for each event.
and found that those for events in two high-count time bins were significantly
shorter than in other time intervals. We further reprocessed the observation data
with a relaxed elapsed-time filter
\citep[see][for more details]{mrh+15} and were able to
recover an additional 58 events in a $R=120''$ circular aperture in the
3--79~keV band for the two time bins combined. From this study,
we find that the two high-count bins were actually connected in time
(i.e., the gap between the two bins was filled by the recovered events)
and the livetime of the detectors was $\sim$1/300 of the exposure.
We also investigated other high-count time bins using a relaxed filter,
and were able to recover marginally additional events only
for burst 6 ($10\pm8$ events), the other GBM detected burst. Below,
results for burst properties are obtained with the data processed
with the relaxed filter.

Although it was reported that the {\it Fermi}-detected bursts are likely
from the magnetar 1E~1841$-$045 \citep[][]{paf+14},
the localization was not unambiguous. In order to localize the bursts
better and see if they are really produced by the magnetar
1E~1841$-$045, we used the {\it NuSTAR} data to measure the position of
bursts 2, 3, 5, and 6, which had sufficiently many counts for such
an analysis. We projected their 3--79~keV event distributions collected
for 2~s onto R.A. and decl. axes, and fit the projected profiles with
a Gaussian plus constant function. 
The results for the burst location offsets from the 1E~1841$-$045
position were $\Delta R = 9''\pm2''$, $8''\pm2''$, $9''\pm3''$, and $5''\pm2''$
for bursts 2, 3, 5 and 6, respectively, where the quoted uncertainties are
purely statistical (1$\sigma$).
Note that aspect reconstruction accuracy of {\it NuSTAR}
is $8''$ (90\%), and so the measured positions are consistent with that
of the magnetar 1E~1841$-$045.

\begin{figure*}
\centering
\begin{tabular}{ccc}
\hspace{-6.0 mm}
\includegraphics[width=2.5 in]{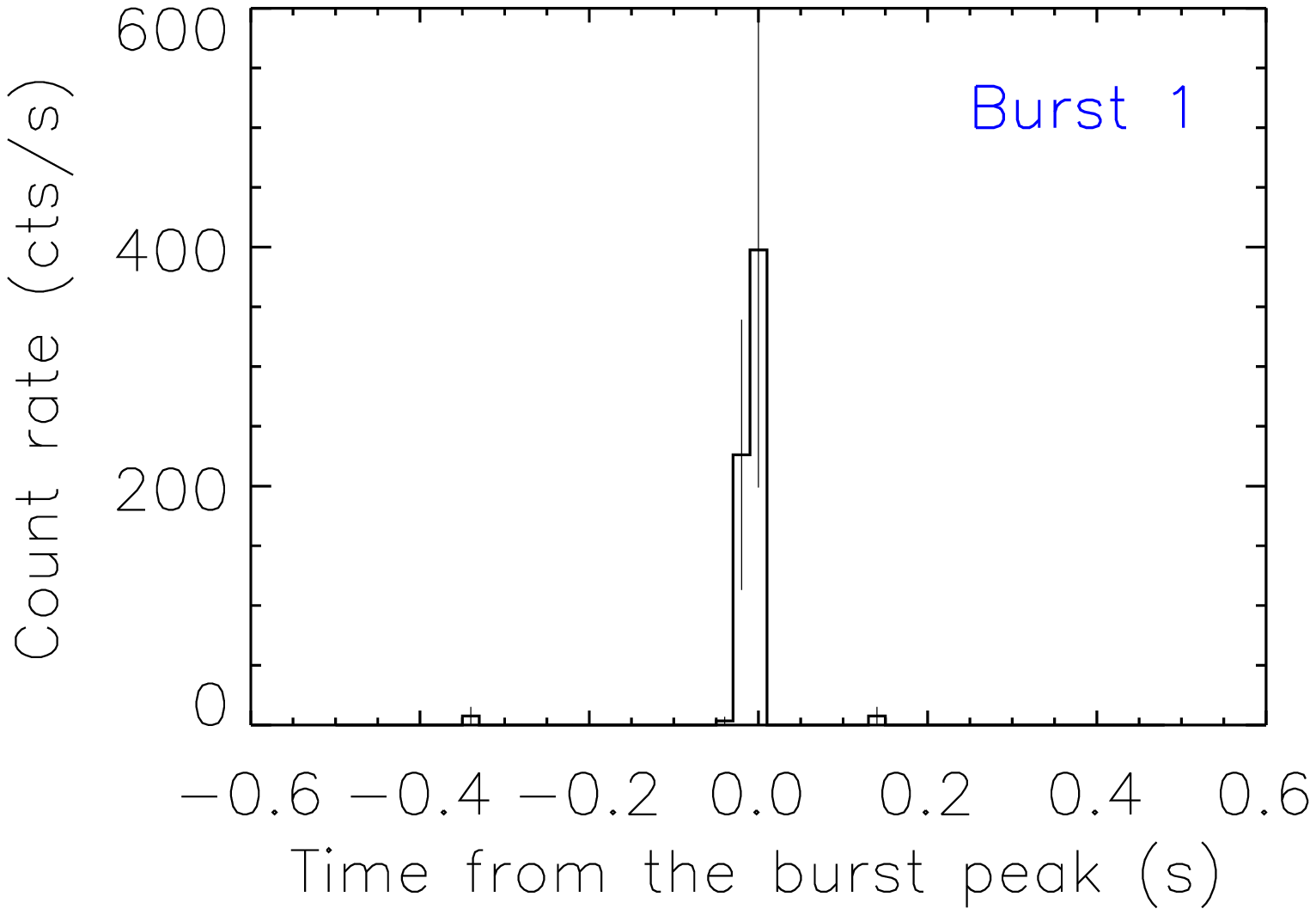} &
\hspace{-7.0 mm}
\includegraphics[width=2.5 in]{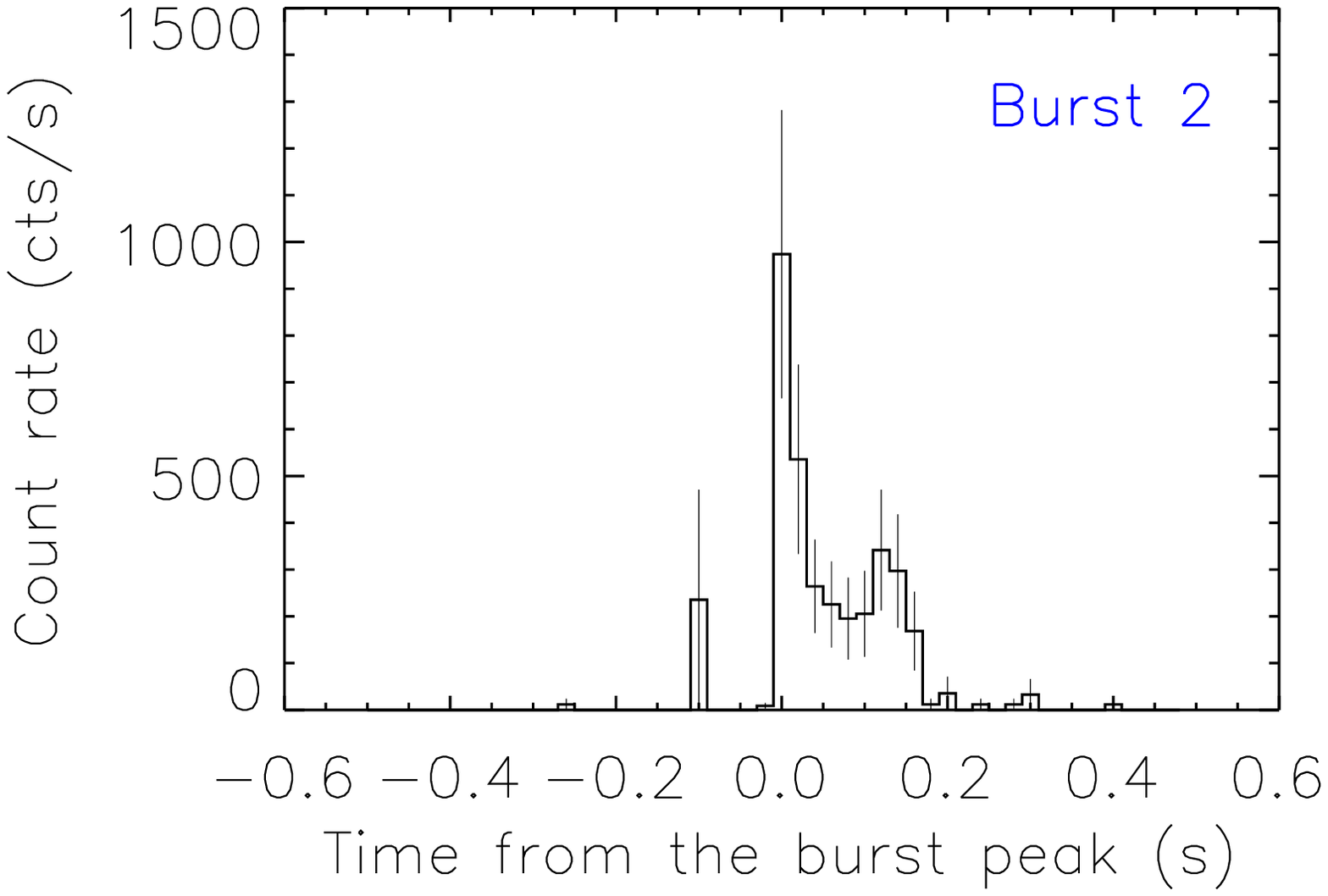} &
\hspace{-7.0 mm}
\includegraphics[width=2.5 in]{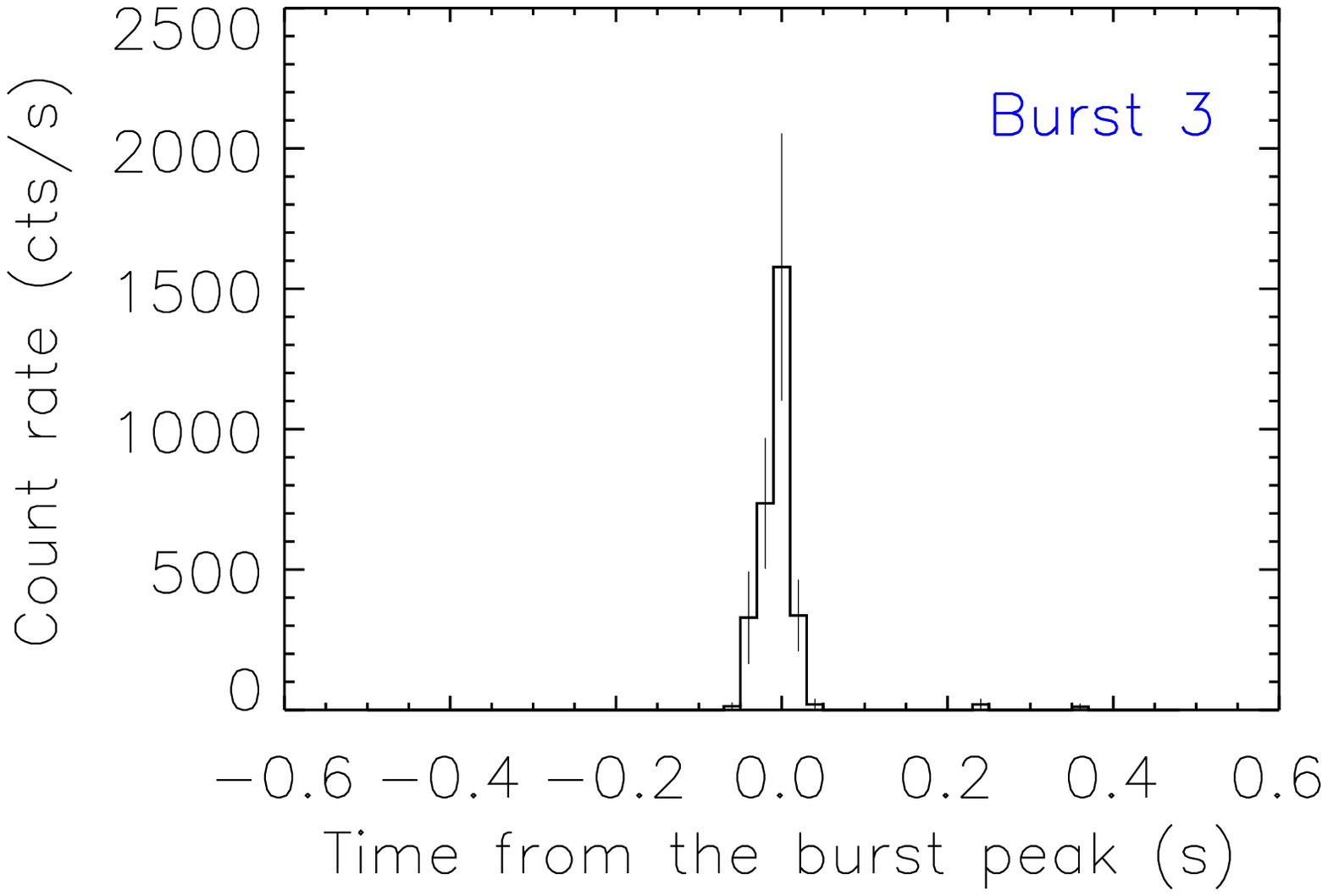} \\
\hspace{-6.0 mm}
\includegraphics[width=2.5 in]{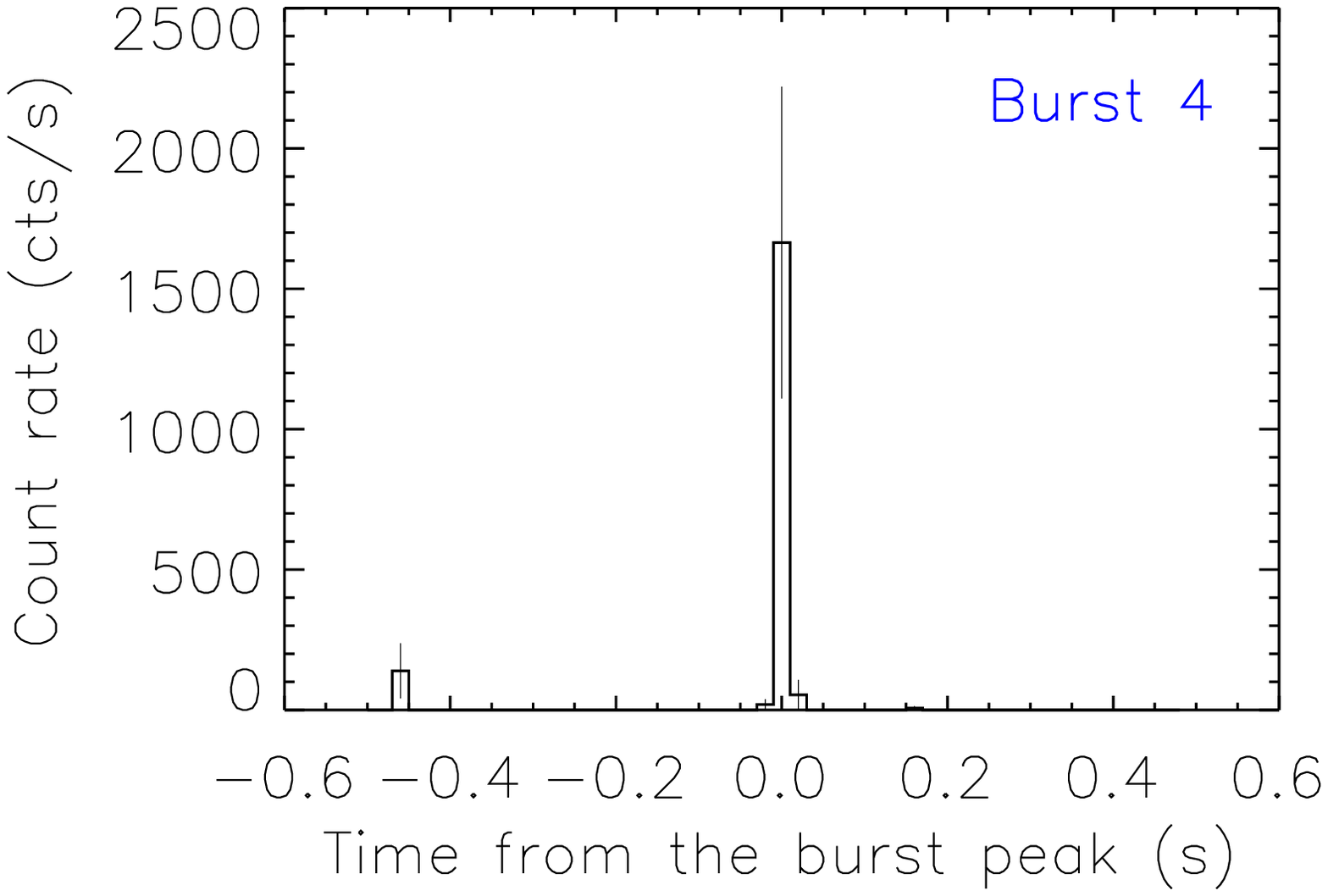} &
\hspace{-7.0 mm}
\includegraphics[width=2.5 in]{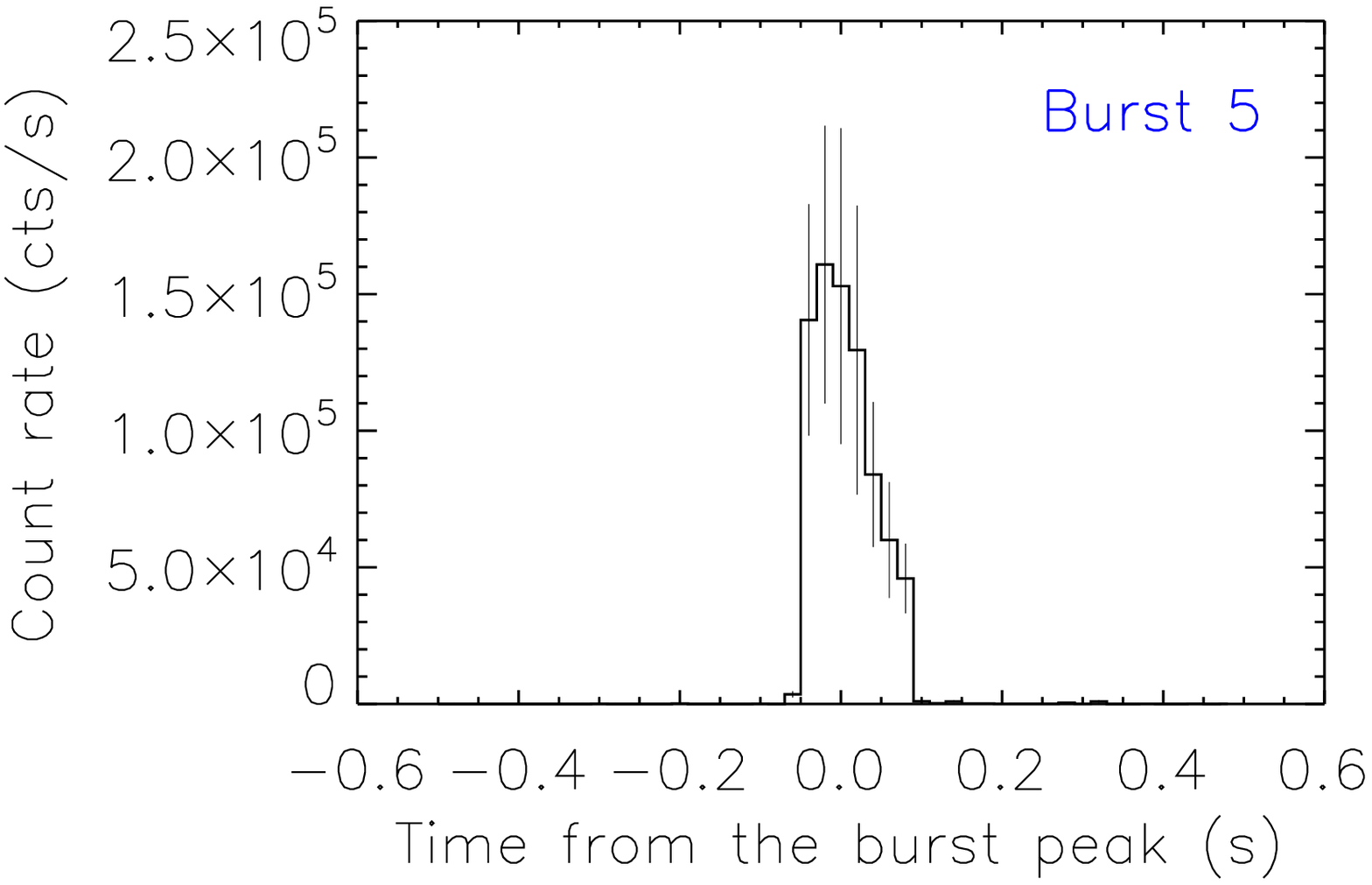} &
\hspace{-7.0 mm}
\includegraphics[width=2.5 in]{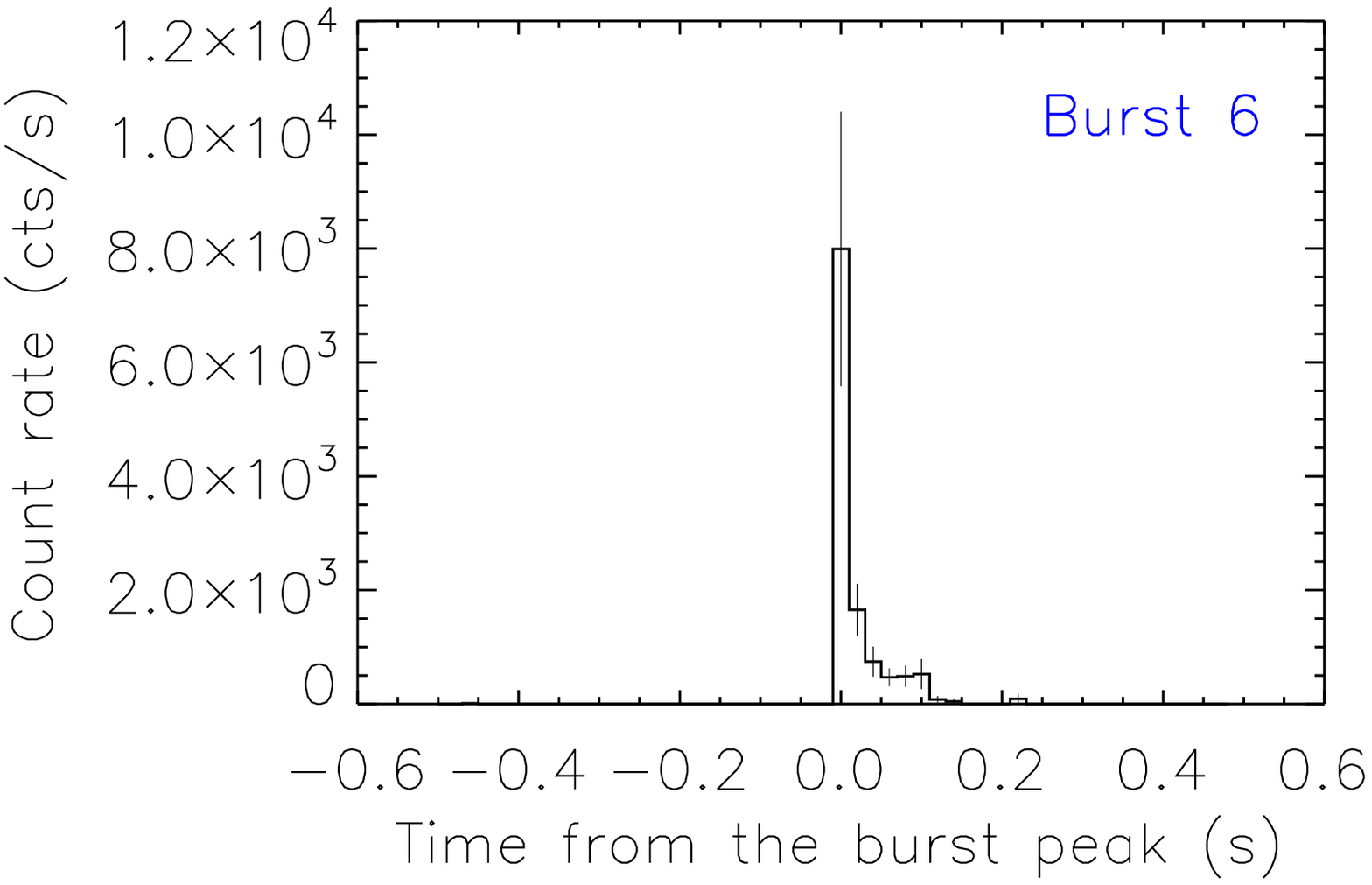} \\
\end{tabular}
\figcaption{Deadtime-corrected light curves of the bursts in the 3--79~keV band.
\label{fig:burstmorp}
}
\vspace{0mm}
\end{figure*}

In order to characterize the properties of the bursts, we fit the short-term
($\sim$10-s) light curves around the bursts with exponentially rising and
falling functions
\begin{equation}
\label{eqn:lightcurve}
F(t) =
 \begin{cases}
   A e^{(t-T_0)/T_{\rm r}} + C_1 &  t<T_0, \\
   (A - C_2) e^{-(t-T_0)/T_{\rm f}} + C_1 + C_2 & t\geq T_0,
 \end{cases}
\end{equation}
where $A$ is the amplitude, $T_{\rm 0}$ is the burst peak time,
$T_{\rm r}$ is the rising time, $T_{\rm f}$ is the falling time,
and $C_{1,2}$ are constants \citep[see][for a different model]{gdk11}.
Since there are only 10--120 counts detected within a 1-s window
around each burst, we extracted events from the whole detector and
used a maximum likelihood optimization without binning.
The peak count rates are very high and the $T_{\rm 90}$ durations
are $\lapp$1 sec. We present the results in Table~\ref{ta:bursts} and
show the bursts morphologies in Figure~\ref{fig:burstmorp}.
Some bursts occurred within 1~sec of each other.
Note that we do not find a clear increase in the tail emission except
for in burst 5.

The spectrum of a burst can provide information on the burst mechanism.
Therefore, we extracted $\sim$0.2-s spectra in circles with radius $R = 120''$
centered at the source position for bursts 2 and 5, and
fit the spectra with single component models;
we tried both a blackbody and a power-law model.
In order to remove the persistent emission, we extracted
background counts
in the same region as we used for the source but in a 1-ks
pre-burst interval. We did not attempt to fit the burst data with
a multi-component model because there were too few photons collected
during the 0.2-s intervals.
We grouped the spectrum to have 1 count per spectral bin, and used
$lstat$ \citep{l92} in {\tt XSPEC} 12.8.1g.
The {\it NuSTAR} bandpass is not sensitive to the relatively low
$N_{\rm H}$ of the source,
and we set it to $2.05 \times 10^{22}\rm \ cm^{-2}$ which we obtained
by joint-fitting of the soft-band spectra with a broken power-law model
(see Section~\ref{sec:phavspec}). We use this value
and the {\tt tbabs} absorption model in {\it XSPEC} throughout this paper
unless noted otherwise. The burst spectra can be described with
a power-law model having $\Gamma =1-2$ or a blackbody model with
$kT =3-5\rm \ keV$.
The results of the fits are shown in Table~\ref{ta:burst90}.
Note that we did not detect the high temperature blackbody
component \citep[$kT = 13$~keV][]{cxc14},
which is probably because of the low statistics at high energy.

\newcommand{\markax}{\tablenotemark{a}}
\newcommand{\markbx}{\tablenotemark{b}}
\begin{table}[t]
\vspace{0.0mm}
\begin{center}
\caption{
Best-fit parameters of the burst spectra for a $\sim$0.2~s
interval around the burst peak}
\vspace{-1.5mm}
\label{ta:burst90}
\scriptsize{
\begin{tabular}{cccc} \hline\hline
Burst & $\Gamma$ & 3--79 flux\markax & $lstat$/dof  \\ 
      &  &   $10^{-8}\ \rm erg\ s^{-1}\ cm^{-2} $    \\ \hline
2     & 1.6(3) & 3.5(1.2) &  37/46   \\ 
5     & 0.96(24) & 1300(600) & 53/56  \\  \hline
Burst & $kT$ & $L_{\rm BB}^\markbx$ & $lstat$/dof \\
      & (keV) & $10^{38}\ \rm erg\ s^{-1}$ &	 \\ \hline
2     & 3.3(3) & 1.5(3) & 48/46    \\
5     & 4.8(5) & 400(100) & 59/56  \\ \hline
\end{tabular}}
\end{center}
\vspace{-2.5mm}
$^{\rm a}$ Deadtime-corrected flux.\\
$^{\rm b}$ Deadtime-corrected bolometric luminosity for an assumed distance of 8.5 kpc.\\
\vspace{0.0mm}
\end{table}

\subsubsection{Burst Tails}

\begin{figure*}
\centering
\begin{tabular}{cc}
\hspace{-3.0 mm}
\includegraphics[width=2.8 in, angle=90]{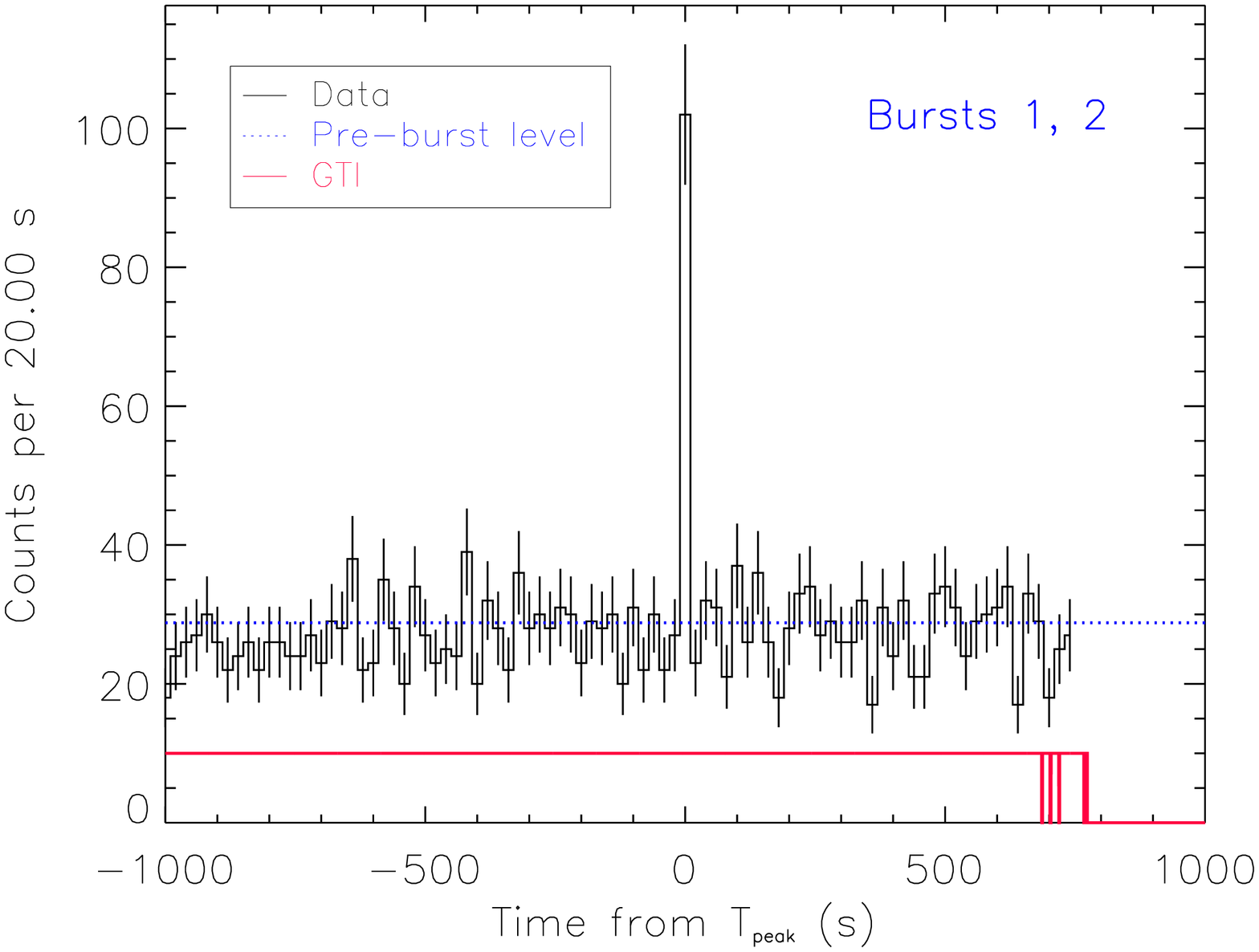} &
\hspace{-13.0 mm}
\includegraphics[width=2.8 in, angle=90]{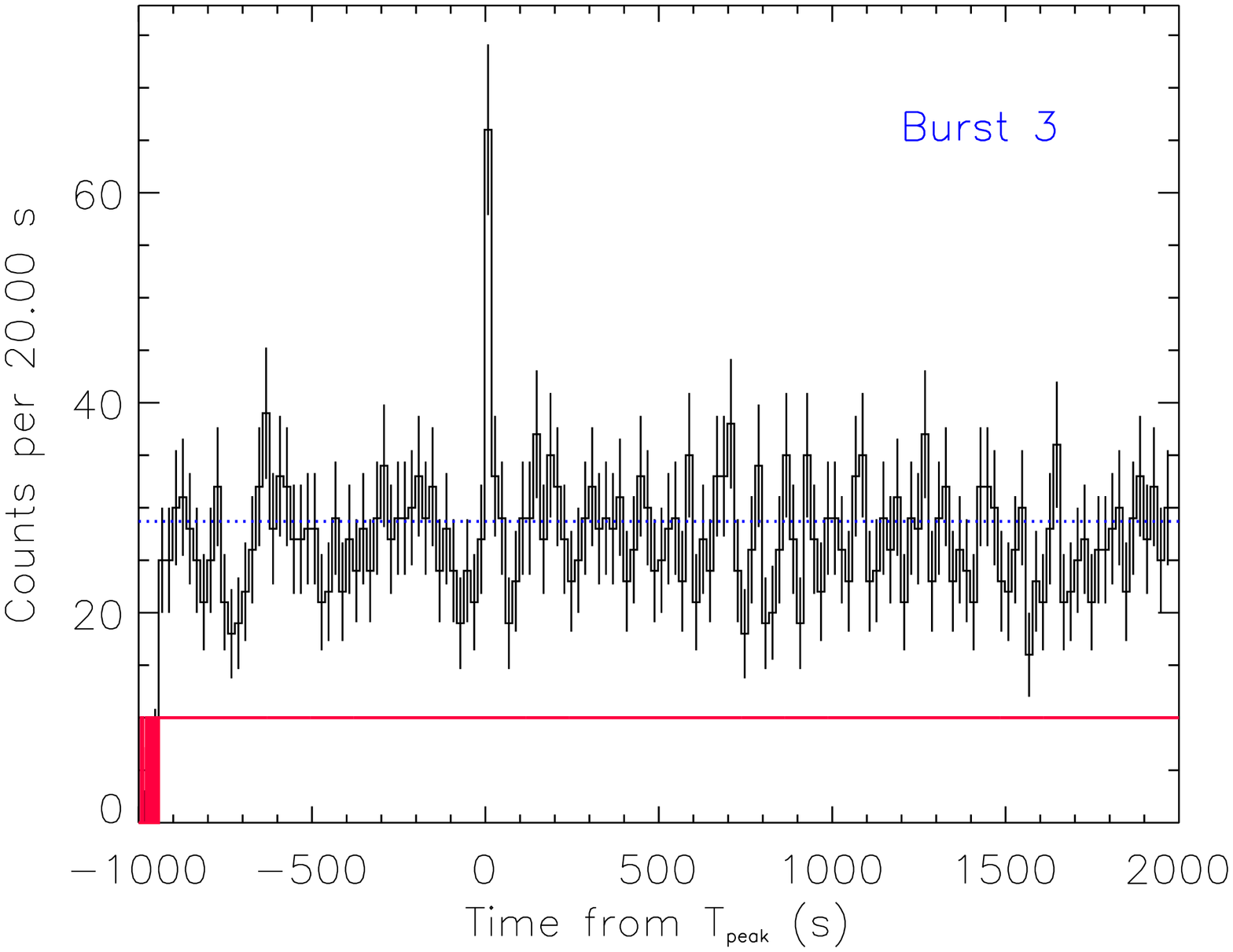} \\
\hspace{-7.0 mm}
\includegraphics[width=2.8 in, angle=90]{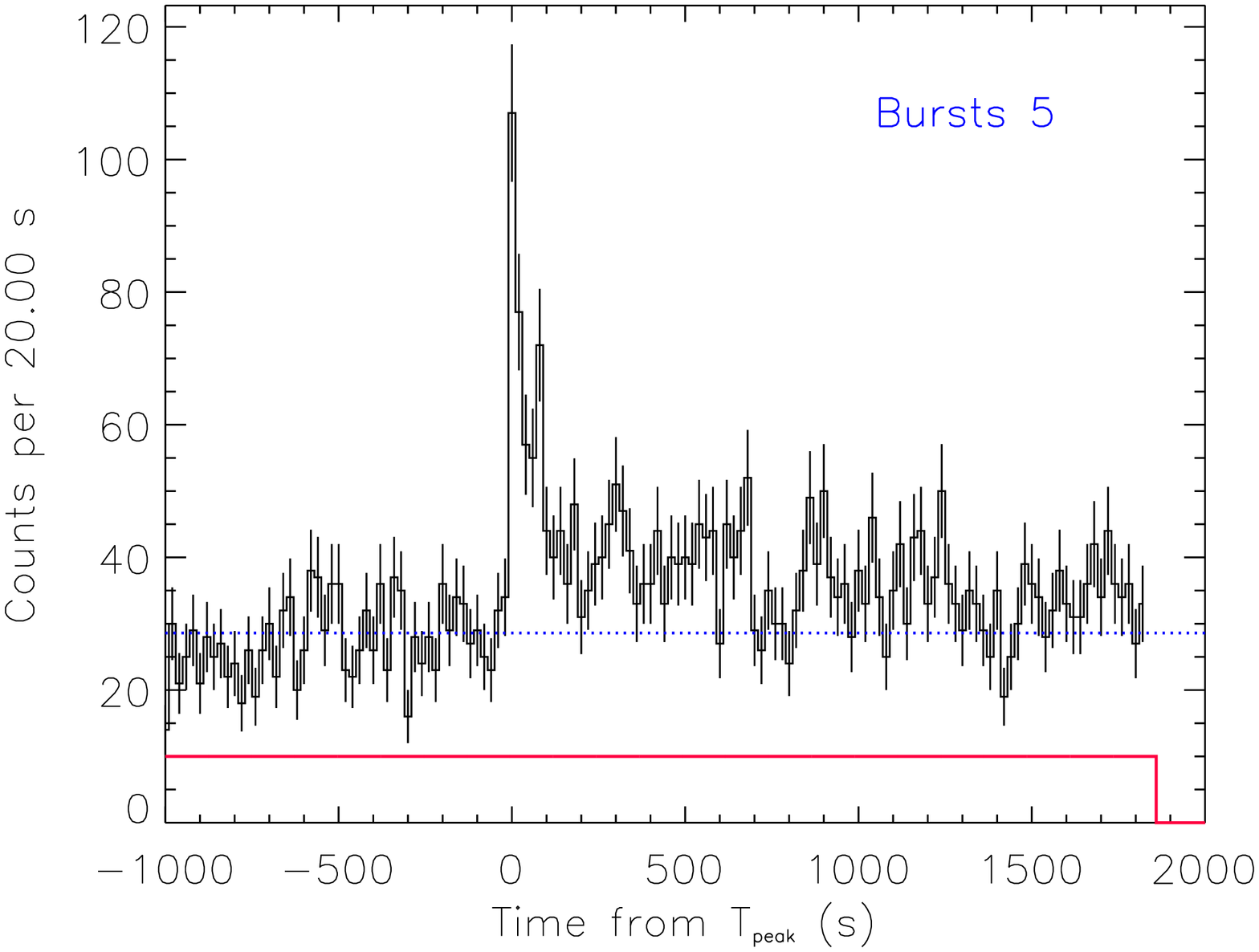} &
\hspace{-13.0 mm}
\includegraphics[width=2.8 in, angle=90]{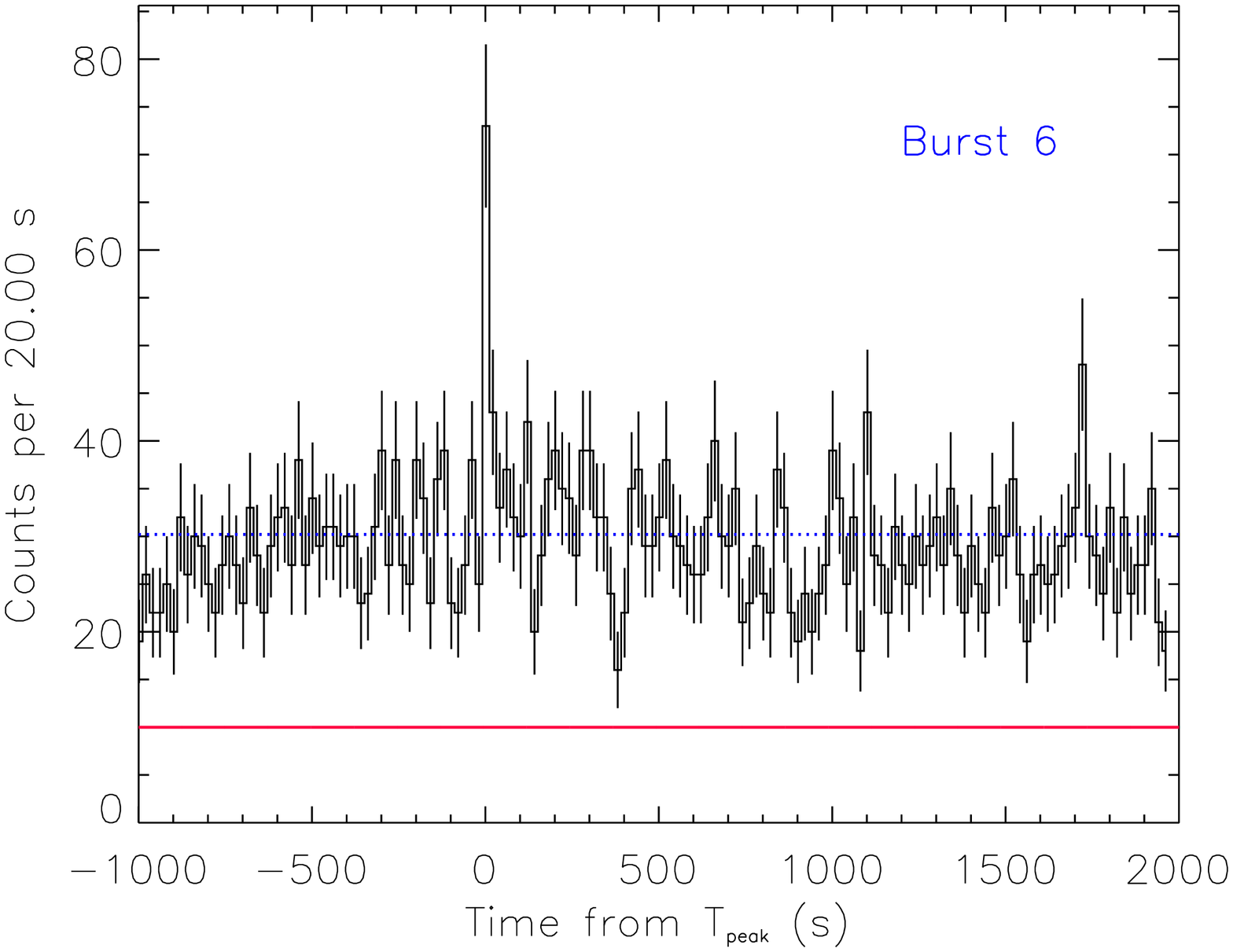} \\
\end{tabular}
\figcaption{Observed long-term light curves of the bursts in the 3--79~keV band.
Note that bursts 1 and 2 are very close in time ($\lapp$1 s), and
that the light curve for burst 4 is not shown because
the burst was not detected in a 20-s time bin. The blue dotted line shows
the pre-burst emission level, and the red line shows the scaled
Good Time Interval (GTI).
\label{fig:burstmorplong}
}
\vspace{0mm}
\end{figure*}

Short X-ray bursts from magnetars can exhibit
long emission tails, lasting for hours \citep[][]{akb+14}.
In order to search for long tails, we binned the light curves
into 20-s bins, and compared counts in 10 bins before and after a burst,
excluding the burst bin. We then compared the pre- and post-burst counts,
and found that the difference is significant only for burst 5
($\Delta_{\rm post-pre} = 211 \pm 43$ counts for a 200-s time interval).
We show the light curves for the bursts in Figure~\ref{fig:burstmorplong},
and report the results in Table~\ref{ta:bursts}.
We performed the same study on different timescales (e.g., 2~s and 50~s),
and found the same results; the difference is significant only for burst 5

\begin{figure*}[h]
\centering
\begin{tabular}{cc}
\hspace{-3.0 mm}
\includegraphics[width=3.5 in]{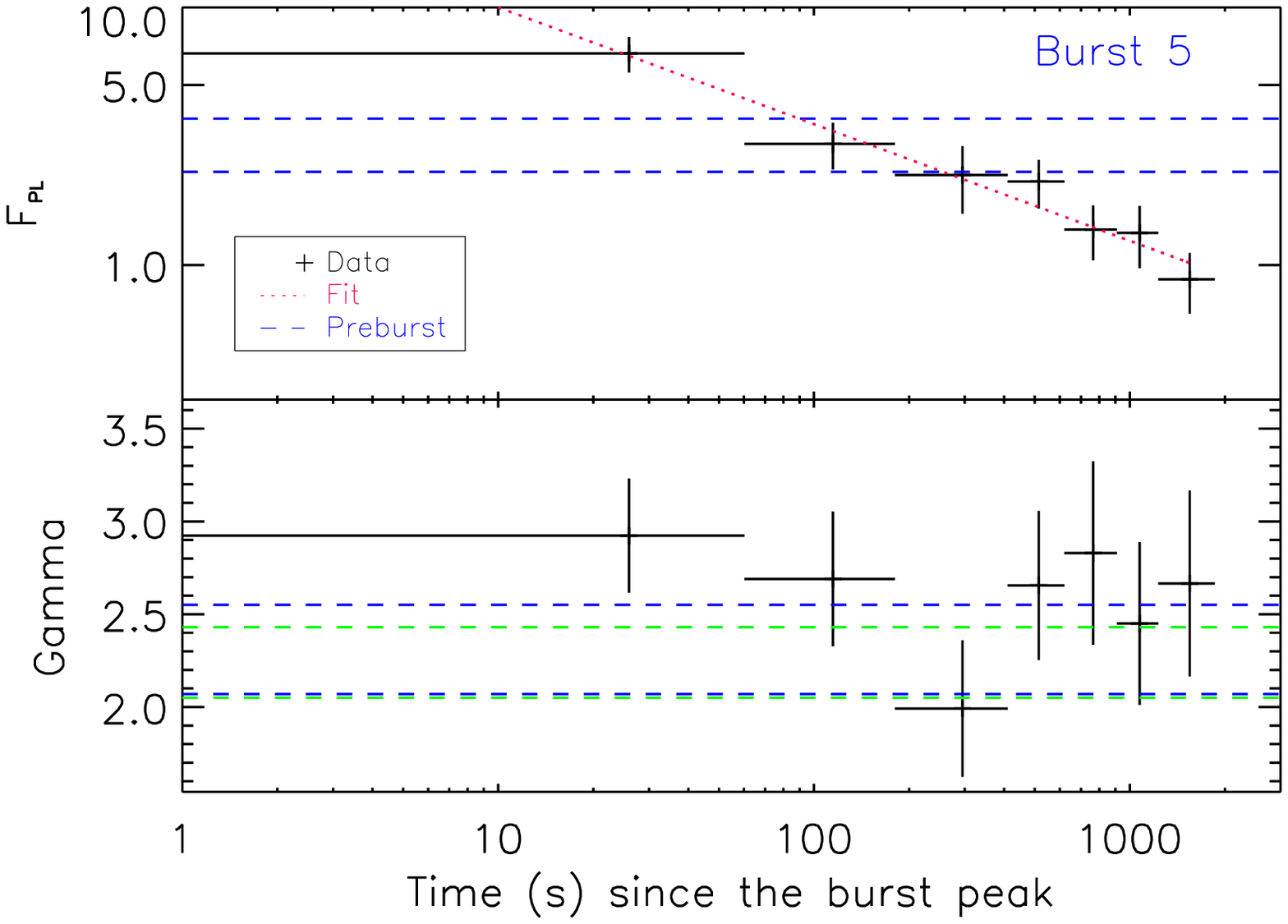} &
\hspace{5.0 mm}
\includegraphics[width=2.4 in, angle=90]{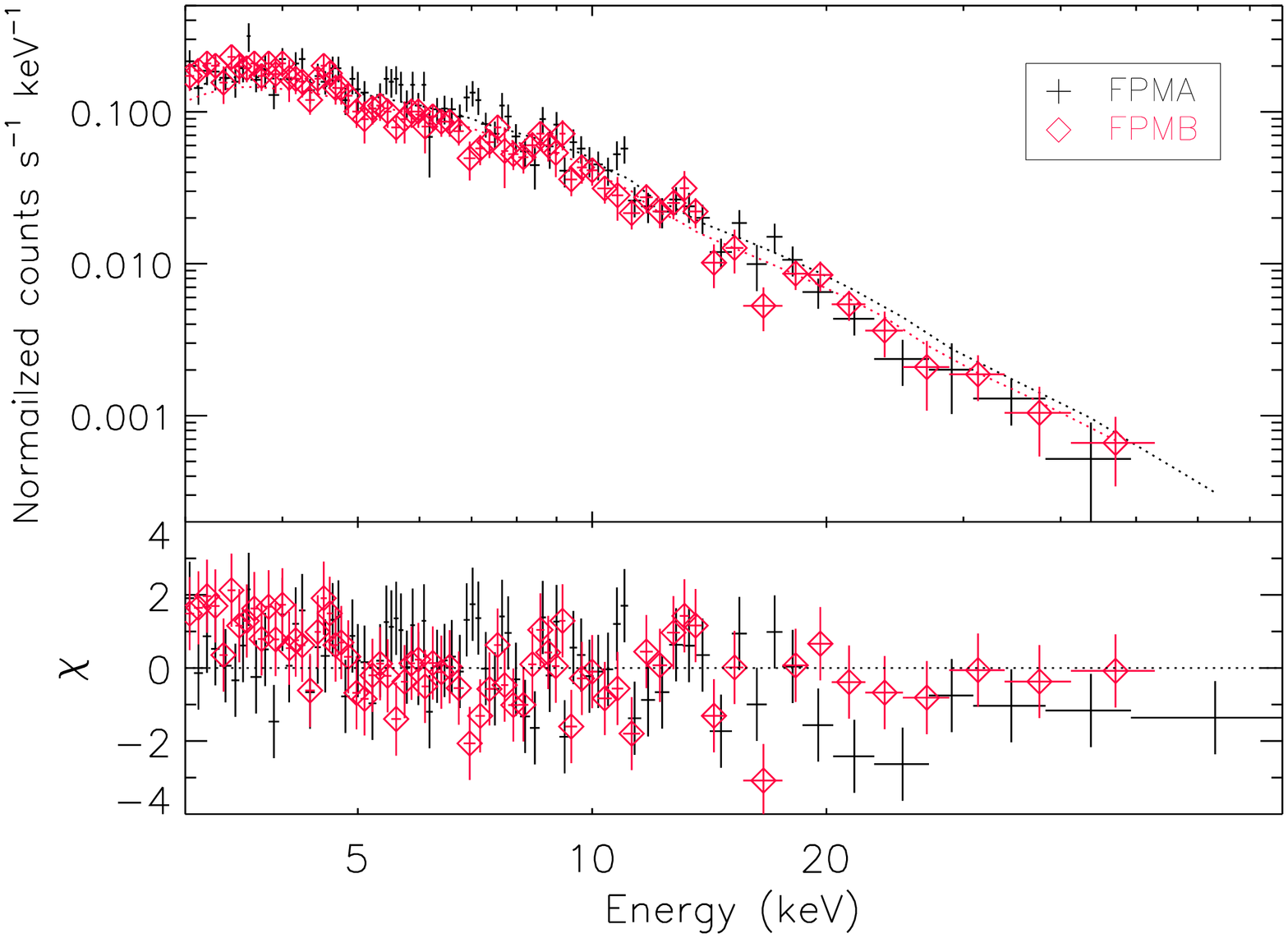}
\end{tabular}
\figcaption{Evolution of the persistent-emission-removed spectrum of the tail
of burst 5 measured using a power-law model (left) and the
integrated spectra (right). We used events collected after $T=T_{\rm 0}+0.5$\,s
to remove the burst emission.
Flux is in units of $10^{-11}\rm \ erg\ s^{-1}\ cm^{-2}$ in the 3--79\,keV band.
The red dotted line in the left panel shows the power-law function
that best describes the flux evolution, blue dashed lines show
the 1$\sigma$ range of the quantities measured in a 100-s
interval before the burst.
The green dashed lines in the left bottom show the 1$\sigma$ range of
the best-fit power-law index for the integrated tail spectrum.
The integrated tail spectrum with a blackbody plus broken power-law
fit with spectral shape parameters frozen at the values in Table~\ref{ta:spec}
is shown in the right panel.
\label{fig:tailevol}
}
\vspace{0mm}
\end{figure*}

In order to search for spectral evolution after burst 5,
we extracted spectra within a radius $R = 120''$ in the tail of
burst 5 in several time intervals excluding the burst
($T>T_{\rm 0}+0.5$\,s; see Table~\ref{ta:bursts}).
Each interval had $\gapp100$ events above the persistent emission
plus background. We then fit the spectra with a blackbody and
a power-law model. We show the results of the power-law fit in Figure~\ref{fig:tailevol}.
The spectral shape did not change significantly over $\sim$2-ks
of tail emission, and the 3--79~keV flux decay is well described
with a power-law function, having a decay index of $0.45\pm0.10$.
Note that the decay index was measured after taking into account
the covariance between the flux and the spectral index in the fit.
A similar decay index is measured for the luminosity
when using a blackbody model. Note that \citet{akb+14} measured
the flux evolution index in the tail to be 0.8--0.9 for 1E~1048.1$-$5937,
much steeper than our measurement for 1E~1841$-$045.
The count enhancement at later times seems to be spiky,
and might have been caused by undetected activity there.
Since the other bursts do not have significant tail emission,
we were not able to measure their spectral evolution.

The tail spectrum of burst 5 is soft compared to the burst spectrum,
and may be similar to the persistent emission.
In order to see if the spectra of the burst tails are significantly
different from the persistent source emission,
we studied the persistent emission in a pre-burst interval in
which approximately the same number of events was collected as
in the tail spectra. We extracted a persistent spectrum
in a 100-s pre-burst time interval, modeled it with a single component model,
either a power law or a blackbody.
The persistent spectrum over the short interval was well described with
single component models, and we find the best-fit parameters are
$\Gamma = 2.31 \pm 0.24$ or $kT = 1.58 \pm 0.13$~keV,
similar to the tail spectrum.
We show the persistent levels with blue dashed lines in Figure 4.

Since the spectral shape did not change significantly over the tail interval,
we fit the combined ($\sim$1.8 ks) tail spectrum to a power-law
or a blackbody model after removing the persistent emission.
The spectrum is well fit by a power-law model
($\chi^2$/dof=122/146) with a photon index $\Gamma=2.2 \pm 0.2$
(green lines in Figure~\ref{fig:tailevol} left),
similar to the 100-s persistent spectrum.
A blackbody model also fits the data
with $kT=2.1\pm0.2\rm \ keV$ ($\chi^2$/dof=137/146).
We tried to fit the combined tail emission with the model
for the persistent spectra obtained below (Section~\ref{sec:phavspec})
using the same fit parameters except for the normalization constants
(Table~\ref{ta:spec}). The model fits the spectrum well
($\chi^2$/dof=171/147 with the null hypothesis probability $p = 0.09$).
However, we see a trend in the fit
residuals which suggests that the tail spectra may be slightly softer
than the persistent one (see Figure~\ref{fig:tailevol} right).

\subsection{Persistent Emission}
In order to study the persistent emission, we removed the burst intervals
using time filters. We used 20-s windows centered at the burst peak times
for all the bursts except for burst 5 for which we used a 2-ks window
because of the tail. For the soft-band spectrum below 10~keV,
we used the {\it Chandra} (only Obs. ID 730 due to pile-up in Obs. ID 6732),
{\it XMM-Newton}, and {\it Swift} data (see Table~\ref{ta:obs}),
the same observations as used by A13.
Although these observations were taken long before the {\it NuSTAR}
observation, we show below that the source emission properties have been
stable over 10 years (see also A13).
For the {\it NuSTAR} data, we used a circular aperture with $R = 60''$
for the source and an annular aperture with inner and outer
radii of $R = 60''$ and $R = 100''$ for the background, respectively.
Note that $\sim$15\% of the source events fall in the background region
because of the {\it NuSTAR} PSF \citep{amw+14}. However, not all the $\sim$15\%
of the source flux is subtracted as background in the spectral fitting
since we scale the area
of the background region to that of the source for spectral fits, and
$\sim$10\% of source events will be lost during background subtraction.
We take into account this effect using a normalization constant.

\subsubsection{Timing Analysis}
\label{sec:timing}
\begin{figure}
\centering
\hspace{-6 mm}
 \includegraphics[width=3.63 in]{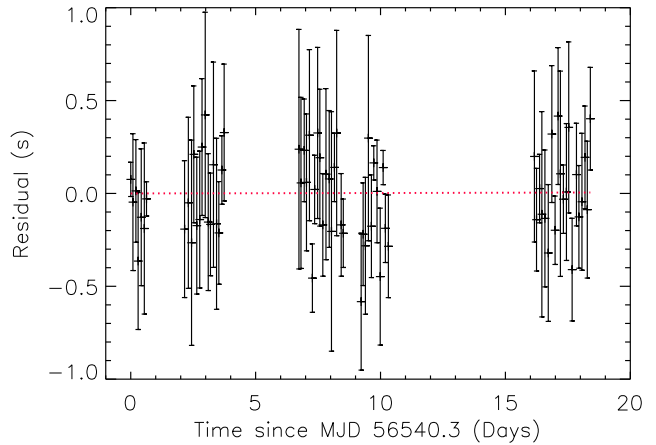}
\vspace{-5 mm}
\figcaption{Timing residuals after fitting the pulse phases
in the 3--79~keV band for observations 30001025004--12.
\label{fig:phasefit}
\vspace{2.0 mm}
}
\end{figure}

For our timing analysis, we extracted source events from the new observations
in the 3--79~keV band within a radius $R = 60''$ of the nominal
source position, and divided the events into subintervals consisting of
$\sim$5000 counts each. Note that we have used different subintervals
and found that the results do not change significantly.
Each subinterval was then folded at the nominal
pulse period to yield pulse profiles each with 64 phase bins.
In order to perform phase-coherent timing, we first cross-correlated
the pulse profiles to measure the phase for each subinterval.
We then fit the phases to a quadratic function using frequency
($\nu$) and its first derivative ($\dot \nu$),
$\phi(t) = \phi_0 + \nu(t - T_{\rm ref}) + \dot\nu(t - T_{\rm ref})^2/2$,
to derive a timing solution. We produced a high signal-to-noise ratio
template by coherently combining the pulse profiles using the timing solution,
and cross-correlated the pulse profiles with the template in order to
refine the timing solution. We show the residuals after the fit
in Figure~\ref{fig:phasefit}. We find that the best timing solution
during the observations has parameters $P = 11.79234(1)\rm \ s$
and $\dot P = 4.2(2) \times 10^{-11}\rm \ s\ s^{-1}$,
implying a magnetic field strength
of $7 \times 10^{14}\rm \ G$. Note that we did not include the
first {\it NuSTAR} observation (Obs. ID 30001025002) in this study
because of phase ambiguity. We verified the timing solution
by measuring the period for the individual observations including
the first {\it NuSTAR} observation (Obs. ID 30001025002) using the
$H$-test \citep{dsr+89}, and by fitting the measured period
to a linear function of period evolution $P(t) = P_0 + \dot P(t-T_{\rm ref})\rm \ s$,
which yields $P = 11.792344(4)\rm \ s$ and
$\dot P = 4.05(9) \times 10^{-11}\rm \ s\ s^{-1}$.
We find that the results of the two methods are consistent
with each other and with the results of our {\it Swift} monitoring
(see Section~\ref{sec:swiftmonitoring}).

\subsubsection{Pulse Profiles and Pulsed Fraction}
\label{sec:PPPF}
The pulse profile of 1E~1841$-$045 is known to change with energy
\citep[][]{khh+06, ahk+13}. In particular, A13 found
that the pulse shape in the 24--35~keV band is different from
those in the adjacent energy ranges, which suggested the existence
of a spectral feature. However, no firm conclusion could be made
due to limited statistics. We investigate this here
with much better statistics.

\begin{figure*}
\centering
\hspace{-10.0 mm}
\includegraphics[width=7.3 in]{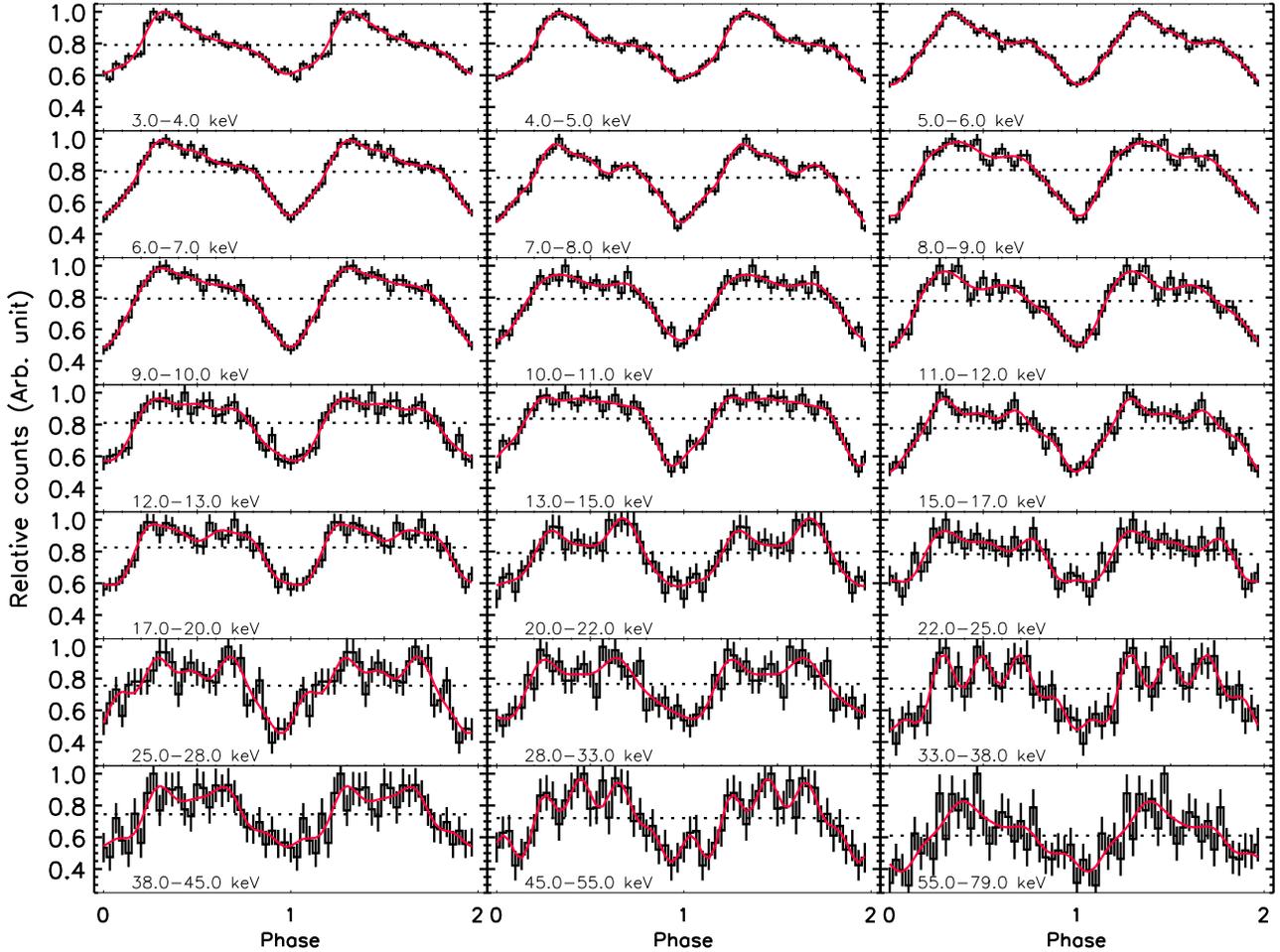} \\
\figcaption{Background-subtracted pulse profiles for 1E~1841$-$045
measured with {\it NuSTAR} in various energy bands. The average value is
shown in a black dotted line and a Fourier reconstructed profile with
five harmonics is shown in red in each panel.
\label{fig:pulseprofile}
}
\vspace{0mm}
\end{figure*}

We produced pulse profiles for individual observations,
and aligned them with the template. Backgrounds were extracted from
an annular region with inner and outer radii of 60$''$ and 100$''$
and were subtracted from the pulse profile of the source region.
We verified that the pulse shape has not changed significantly over
the $\sim$300 days of {\it NuSTAR} observations (Table~\ref{ta:obs})
by comparing the pulse profile of individual observations with
the combined profile in several energy bands (e.g., 3--6~keV,
6--10~keV, 10--15~keV, and 15--79~keV).
We show the combined pulse profiles in several energy bands
in Figure~\ref{fig:pulseprofile}. Note that we find double-peaked
structure similar to that seen by A13, e.g.,
in the 17--33~keV band (see Figure~\ref{fig:pulseprofile}).
However, with the much better statistics we have now,
we find that the pulse shape does not change suddenly but
instead changes gradually with energy. This does not support the existence
of a narrow spectral feature as suggested by A13.

We find that the pulse shape at higher energies becomes more complicated,
sometimes showing three peaks (e.g., 33--38~keV in
Figure~\ref{fig:pulseprofile}; a new peak seems to appear at phase $\sim$0.5).
In order to see if the triple-peaked structure at high energies
($\gapp$ 25~keV) is significant, we fit each pulse profile with
a harmonic function in which the number of harmonics contained was varied
between zero (constant) and three.
In the fit, we calculate the $\chi^2$ value and the F-test probability
by adding higher-order harmonic functions one by one.
From this study, we find that the pulse profiles below 38~keV are
generally well fit with sum of two harmonics,
and the others with a single harmonic function;
adding one more harmonic to these is not statistically required
(99\% confidence). We further fit the pulse profiles with a sum of
the first two harmonics plus a fifth harmonic because the triple-peaked
structure is best described with a fifth harmonic. The
F-test probability for adding the fifth harmonic shows an
improvement to the fit with 99.7\% confidence in the 45--55~keV
band and with 98.6\% confidence in the 33--38~keV band.
However, these may not imply a significant detection when
considering the number of double- peaked profiles we have.
Therefore, we conclude that the double-peaked structures are statistically
significant but the triple-peaked ones are only marginally so.

The pulsed fraction of the source has previously been measured to be
increasing with energy \citep[][A13]{khh+06}.
Note that A13 did not attempt to measure the area pulsed
fraction defined by
\begin{equation}
\label{eq:PFarea}
PF_{\rm area}=\frac{\sum_i (p_i - p_{\rm min})}{\sum_i p_i},
\end{equation}
where $p_i$ is counts in $i$-th phase bin and $p_{\rm min}$ is the
counts from the lowest bin in the pulse profile,
due to insufficient statistics,
and that $PF_{\rm area}$ is known to be biased
upwards in the low counts regime.
Since we now have much better statistics, we measured the area pulsed
fractions in different energy bands and show them in Figure~\ref{fig:pulsedFrac}.
Even with good statistics, the area pulsed fraction is known to be
biased upwards (see Appendix), and so we show two alternative measures
of the pulsed fraction as well. The first alternative is to fit
the pulse profile with a harmonic function, and use the best-fit function
to calculate the pulsed fraction ($PF_{\rm fit}$),
which will remove the bias caused by incorrect identification of
the baseline level by selecting
the minimum phase bin. We find that the pulse profiles are well described
with two harmonics having $\chi^2$/dof$\sim$1 (Figure~\ref{fig:pulseprofile})
except for the lowest energy band for which the two-harmonic function yields
an unacceptable fit, with $p = 10^{-4}$. The large $\chi^2$ in this case is
mostly due to a sharp step in the pulse profile at phase $\sim$0.2,
but the overall pulse shape agrees with the best-fit function well.
Furthermore, rebinning the phase can make the fit acceptable.
Therefore, we used two harmonics for the fit function, calculated
$PF_{\rm area}$ using the best-fit parameters, and show the pulsed
fractions in Figure~\ref{fig:pulsedFrac} (blue squares).
We also calculated the rms pulse amplitude defined by
\begin{equation}
\label{eq:PFrms}
PF_{\rm rms}=\frac{\sqrt{2\sum_{k=1}^{6}((a_k^2+b_k^2)-(\sigma_{a_k}^2+\sigma_{b_k}^2))}}{a_0},
\end{equation}
where
$$a_k=\frac{1}{N}\sum_{i=1}^{N}p_i \cos(\frac{2\pi ki}{N}),
b_k=\frac{1}{N}\sum_{i=1}^{N}p_i \sin(\frac{2\pi ki}{N}),$$
$$\sigma_{a_k}^2 + \sigma_{b_k}^2 =
\frac{1}{N^2}\sum_{i=1}^{N}\sigma_{p_i}^2\cos^2(\frac{2\pi ki}{N}) +
\frac{1}{N^2}\sum_{i=1}^{N}\sigma_{p_i}^2\sin^2(\frac{2\pi ki}{N})$$
is the Fourier power produced by the noise in the data, 
$p_i$ is the number of counts in $i$th bin, $N$ is the total number of bins,
$\sigma_{p_i}$ is uncertainty in $p_i$,
and $n$ is the number of Fourier harmonics included, in this case,
$n=6$ \citep[see][for more details]{abp+14}. We show the results in
Figure~\ref{fig:pulsedFrac} (red diamonds).

\begin{figure}
\vspace{-2.5 mm}
\centering
\hspace{-5.0 mm}
\vspace{-8 mm}
\includegraphics[width=3.5 in]{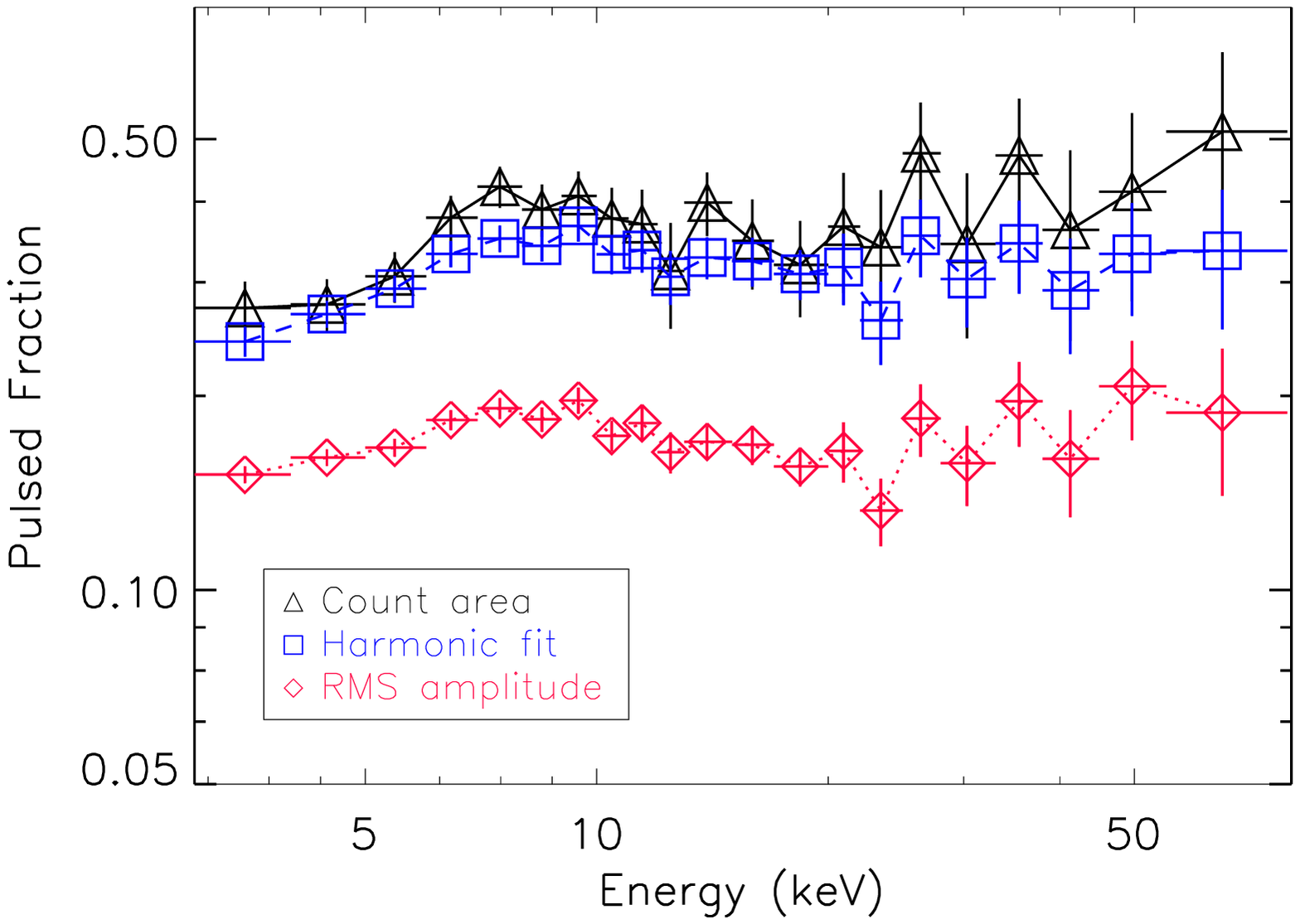} \\
\vspace{5.0 mm}
\figcaption{Pulsed fractions
at several energy bands measured with {\em NuSTAR}.
{\it Black triangles}: area pulsed fraction (Equation~\ref{eq:PFarea});
{\it blue squares}: area pulsed fraction measured using harmonic fit; and
{\it red diamonds}: rms pulsed fraction (Equation~\ref{eq:PFrms}).
\vspace{0.5 mm}
\label{fig:pulsedFrac}
}
\end{figure}

We note that the small-scale features in Figure~\ref{fig:pulsedFrac}
change when we use
different energy resolution. For example, a sudden jump sometimes appears
at $\sim$30~keV similar to that seen by A13 (their Figure 3)
if we use different energy bins. However, the overall trend is similar;
we do not see any rapid increase in the pulsed fraction with energy
above 10\,keV.
This is different from previous reports \citep[][]{khh+06} of pulsed
fraction increasing with energy and approaching 100\% at 100~keV.

\subsubsection{Phase-Averaged Spectral Analysis}
\label{sec:phavspec}

\begin{figure}
\vspace{-2.5 mm}
\centering
\hspace{-5.0 mm}
\vspace{-8 mm}
\includegraphics[width=2.3 in, angle=-90]{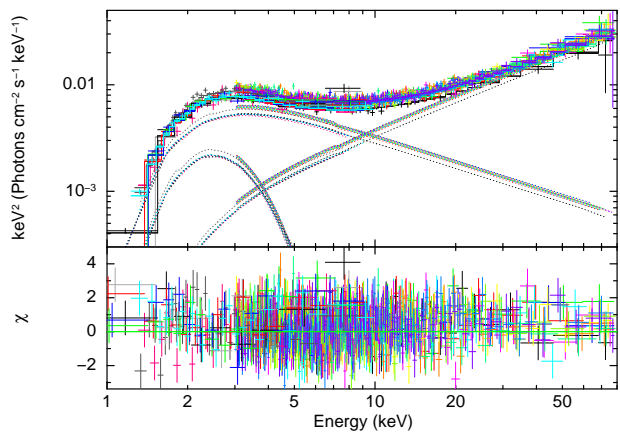}
\vspace{9.0 mm}
\figcaption{Phase averaged spectra of 1E~1841$-$045 and the fit
result. {\it Chandra}, {\it XMM-Newton}, and {\it Swift} data cover below 10~keV,
and {\it NuSTAR} data cover 3--79~keV (see Table~\ref{ta:obs} for observation
summary). Each component of the best-fit model, an absorbed
blackbody plus double power law, is shown in lines. See Table~\ref{ta:spec} for
best-fit model parameters.
\vspace{0.0 mm}
\label{fig:bknspec}
}
\end{figure}

\newcommand{\markt}{\tablenotemark{a}}
\newcommand{\markz}{\tablenotemark{b}}
\newcommand{\markv}{\tablenotemark{c}}
\newcommand{\marku}{\tablenotemark{d}}
\newcommand{\marky}{\tablenotemark{e}}
\newcommand{\markw}{\tablenotemark{f}}
\newcommand{\markx}{\tablenotemark{g}}
\newcommand{\markss}{\tablenotemark{h}}
\begin{table*}
\vspace{0.0 mm}
\begin{center}
\caption{Phenomenological spectral fit results for 1E~1841$-$045
\label{ta:spec}}
\scriptsize{
\begin{tabular}{ccccccccccccc} \hline\hline
Phase & Data\markt& Energy & Model\markz & $N_{\rm H}$ 	& $kT$	&  $\Gamma_{\rm s}\markv$	& $E_{\rm break}/F_{\rm s}$\marku& $\Gamma_{\rm h}$/$\beta$\marky & $F_{\rm h}$\markw & $L_{\rm BB}$\markx & $\chi^2$/dof\\ 
&& (keV) & & ($10^{22}\ \rm cm^{-2}$) 	& (keV)	&		&  (keV/ )&		&  &  &	 \\ \hline
0.0--1.0&N,S,X,C& 0.5--79 & BB+BP	& 2.05(3) & 0.491(5) & 1.95(1) & 13.5(2)/$\cdots$& 1.24(1) & 5.88(6) & 1.64(5) & 8060/7930 \\
0.0--1.0&N,S,X,C& 0.5--79 & BB+2PL      & 2.49(5) & 0.443(9) & 2.82(8) & $\cdots$/1.53(6)& 0.97(4) & 4.70(6) & 1.15(9) & 7931/7930 \\  \hline
Pulsed&N,X,C& 0.5--79 &  PL & 2.05\markss & $\cdots$ & $\cdots$&  $\cdots$    & 1.83(3)& 1.4(1) & $\cdots$   & 1236/1749 \\
Pulsed&N& 3--79 &  PL & 2.05 & $\cdots$             & $\cdots$&  $\cdots$	& 1.81(3) & 1.4(1) & $\cdots$ & 1128/1591 \\
Pulsed&N& 5--79 &  PL & 2.05 & $\cdots$             & $\cdots$&  $\cdots$	& 1.73(4) & 1.4(1) & $\cdots$ & 806/1140 \\
Pulsed&N& 10--79 & PL & 2.05 & $\cdots$             & $\cdots$&  $\cdots$	& 1.47(7) & 1.5(2) & $\cdots$ & 349/510 \\ 
Pulsed&N& 15--79 & PL & 2.05 & $\cdots$             & $\cdots$&  $\cdots$	& 1.28(12)& 1.6(3) & $\cdots$ & 174/279 \\ 
Pulsed&N& 20--79 & PL & 2.05 & $\cdots$             & $\cdots$&  $\cdots$	& 1.0(2)  & 1.5(4) & $\cdots$ & 113/159 \\ \hline
\end{tabular}}
\end{center}
\vspace{-1.0 mm}
\footnotesize{
$^{\rm a}${ N: {\em NuSTAR}, S: {\em Swift}, X: {\em XMM-Newton}, C: {\em Chandra}.}\\
$^{\rm b}${ BB: Blackbody, PL: Power law, BP: Broken power law, and 2PL: Two power laws in {\tt XSPEC}.}\\ 
$^{\rm c}${ Photon index for the soft power-law component.}\\
$^{\rm d}${ Break energy for the BB+BP fit or soft power-law flux
in units of $10^{-11}\ \rm erg\ s^{-1}\ cm^{-2}$ in the 3--79~keV for the BB+2PL fit.}\\
$^{\rm e}${ Photon index for the hard power-law component.} \\
$^{\rm f}${ Flux in units of $10^{-11}\rm \ erg\ cm^{-2}\ s^{-1}$.
The values are only the power-law (hard power-law) flux in the 3--79~keV
band for the BP (PL, 2PL) model.}\\
$^{\rm g}${ Blackbody luminosity in units of $10^{35}\ \rm erg\ s^{-1}$
for an assumed distance of 8.5 kpc \citep[][]{tl08}.}\\
$^{\rm h}${ $N_{\rm H}$ for the pulsed spectral analysis was frozen.}\\}
\vspace{-0.1in}
\end{table*}

Since it is possible that the source has different emission
properties during the bursting periods, we compared
the spectral properties of individual observations. In this
study, we did not include the soft-band spectra because
they were taken at much earlier epochs. We jointly fit all
the {\it NuSTAR} spectra in the 6--79~keV band with an absorbed
broken power-law model in order to minimize the effect
of the blackbody component, which is negligible above 6\,keV,
and found that the spectral shapes for the six {\it NuSTAR}
observations (see Table~\ref{ta:obs}) are
consistent with one another.

Since there is no clear evidence that the source spectral
shape has varied during the {\it NuSTAR} observations,
we used all the {\it NuSTAR} observations for the phase-averaged
spectral analysis. Furthermore, we used the
soft-band data as well in this study since the shape of
the soft-band spectrum is also known to be stable \citep[][]{zk10,dk14}.
We tied all the model
parameters between {\it NuSTAR}, {\it Swift}, {\it XMM-Newton}, and
{\it Chandra} except for the cross-normalization factors. The
normalization constant for {\it NuSTAR} FPMA (Obs. ID
30001025002) was set to be 0.9 as a reference in order to
account for the source contamination in the background
region (Section~\ref{sec:PPPF}). To fit the data, we grouped the
spectra to have at least 20 counts per bin. We used an
absorbed blackbody plus broken power law, and an absorbed
blackbody plus a double power law to fit the 3--79~keV {\it NuSTAR}
data and the 0.5--10~keV soft-band data. We present the
results in Table~\ref{ta:spec} and the spectra in Figure~\ref{fig:bknspec}.

We note that the spectral parameters we report in Table~\ref{ta:spec}
are slightly different from those reported previously (A13).
In order to see if the difference is due
to the updated calibration, we analyzed the same data
set that A13 used (Obs. ID 30001025002),
and were able to reproduce their results except for the
flux. The flux we measure is lower by $\sim$15\% than what
A13 reported, because of a $\sim$15\% increase
in the {\it NuSTAR} effective area from
CALDB 20131007.\footnote{http://heasarc.gsfc.nasa.gov/docs/heasarc/caldb/nustar/docs\\/release\_20131007.txt}
The hard-band component is much better constrained
with the new long exposures, and thus the new results
we report are more accurate. We note that the new parameters
in Table~\ref{ta:spec} are not inconsistent with the data
A13 used, providing acceptable fits to the data
with $\chi^2$/dof =2965/2878 and 2914/2878 for the blackbody
plus broken power law and the blackbody plus double power law, respectively.

\subsection{Phase-Resolved and Pulsed Spectral Analyses}
\label{sec:phresspec}

\begin{figure*}
\centering
\begin{tabular}{ccc}
\hspace{12 mm}
\vspace{0 mm}
\includegraphics[width=1.45 in, angle=90]{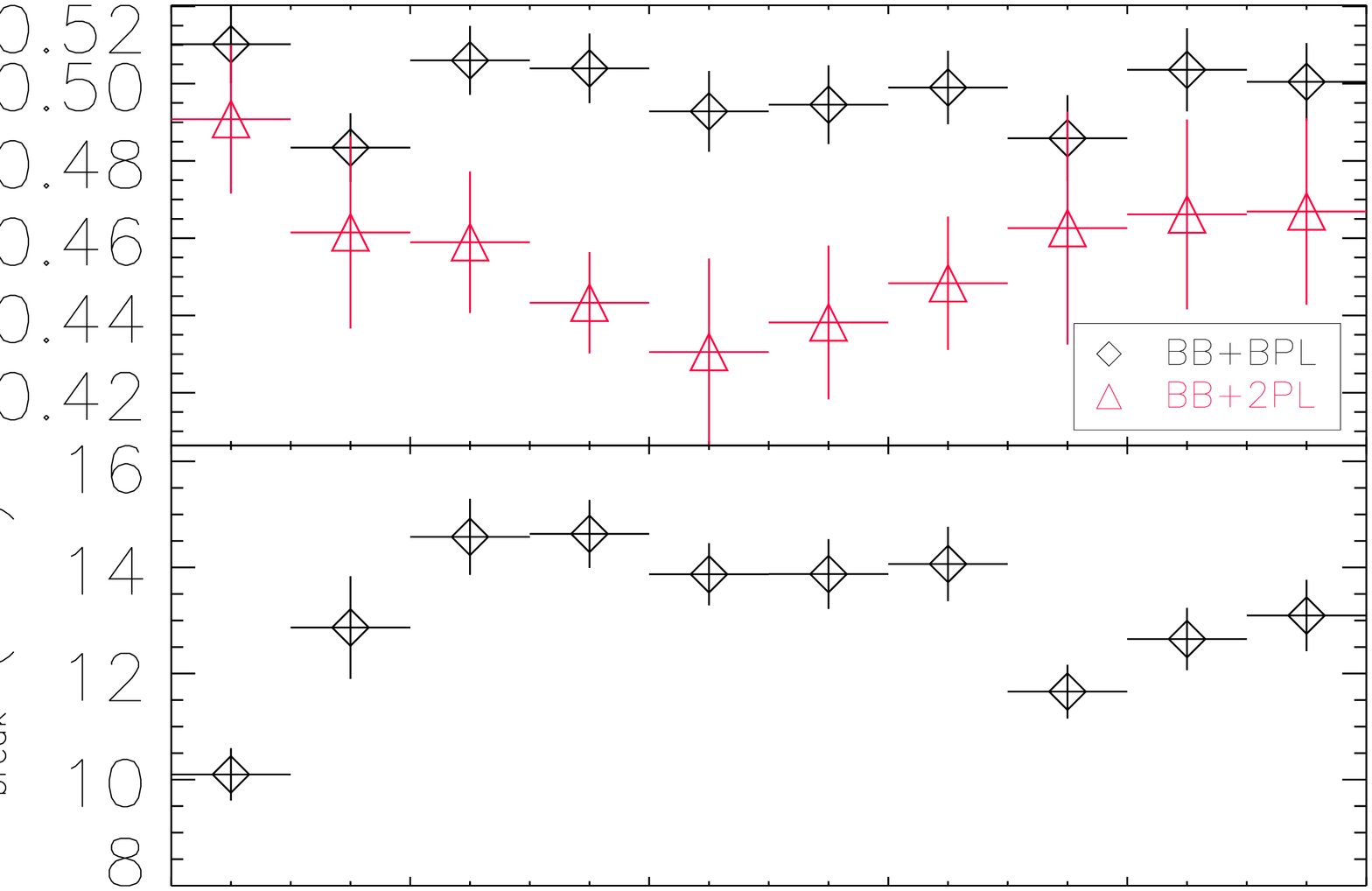} &
\hspace{3.0 mm}
\includegraphics[width=1.45 in, angle=90]{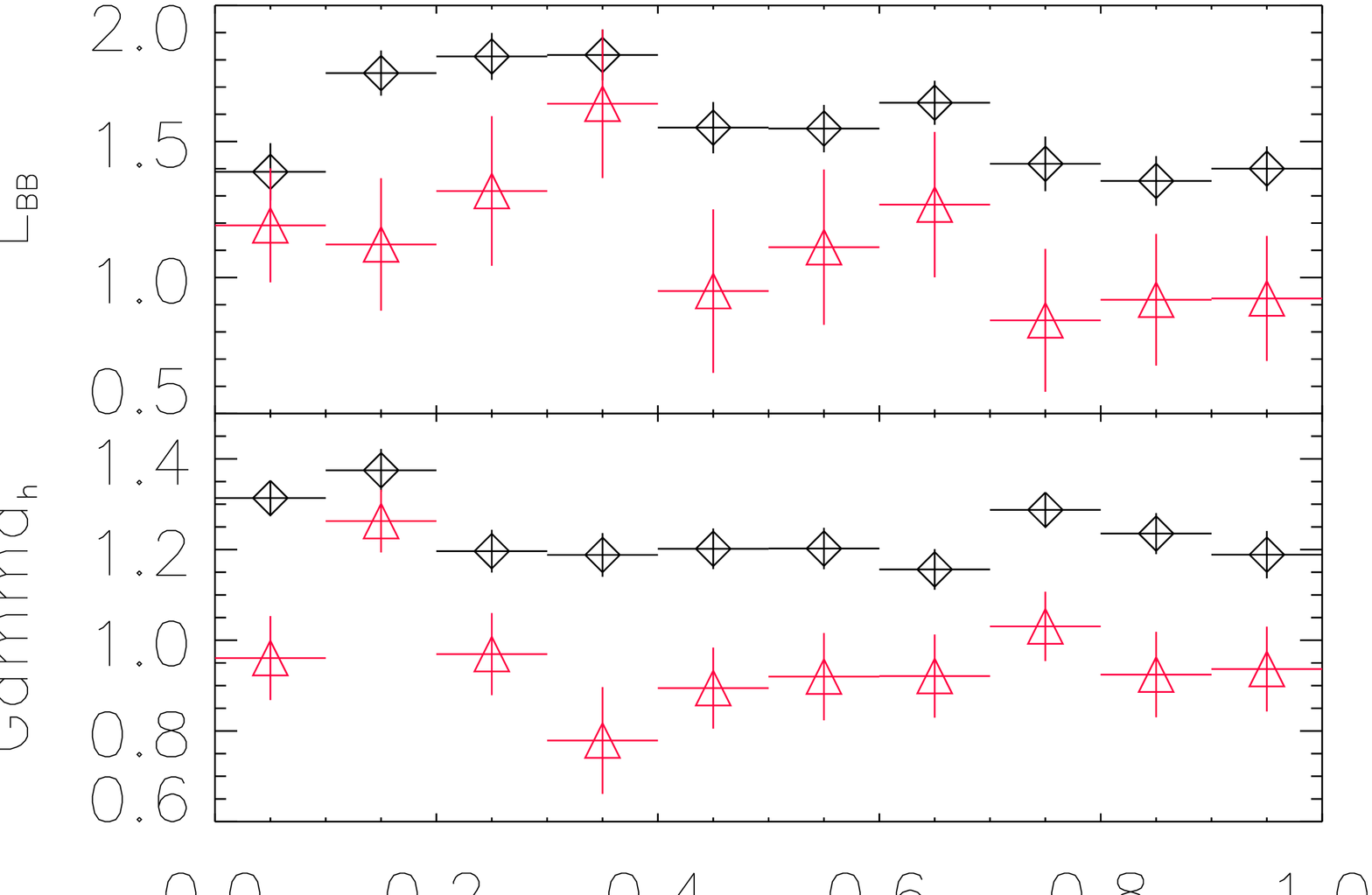} &
\hspace{5.5 mm}
\includegraphics[width=1.45 in, angle=90]{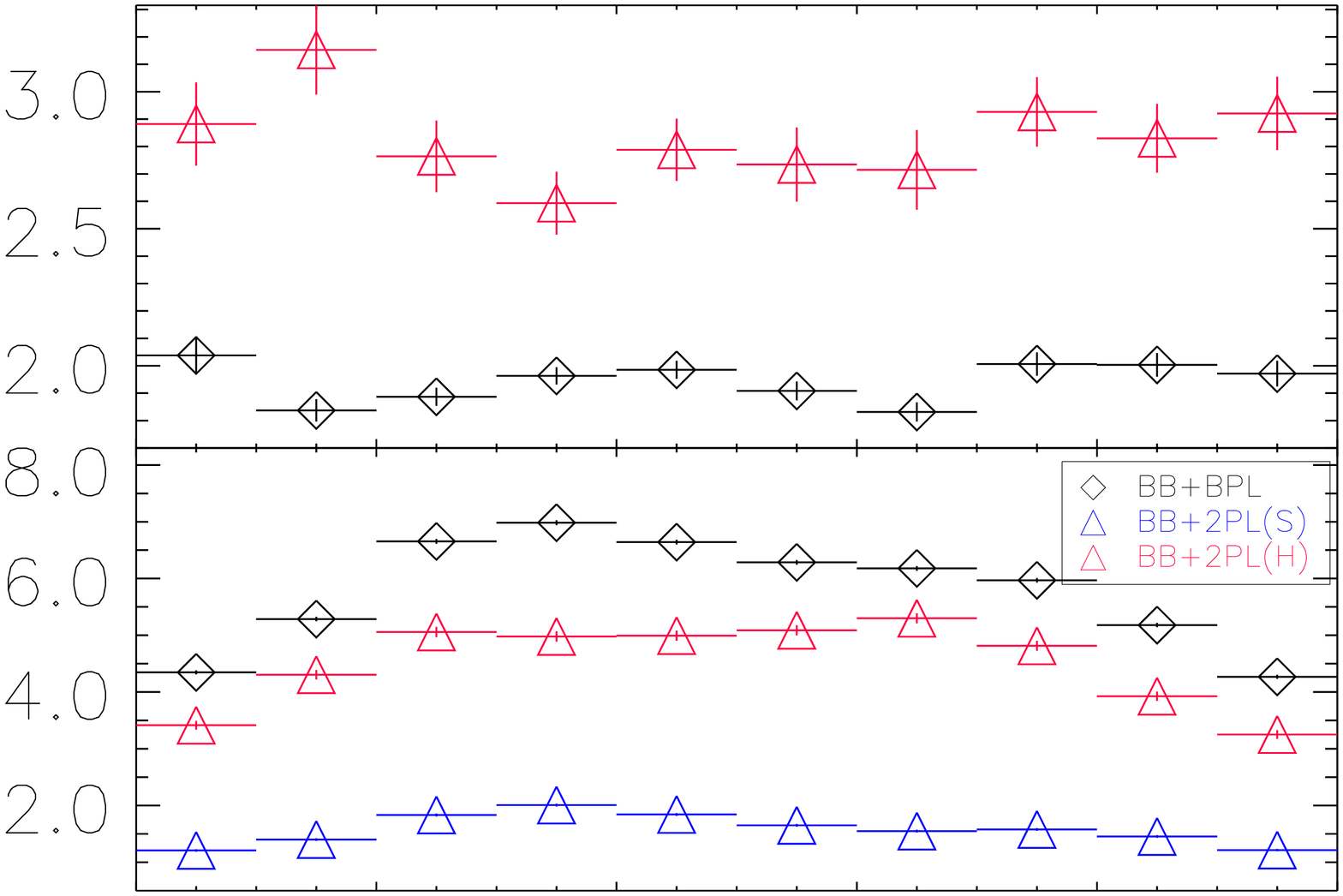} \\
\vspace{1 mm}
\end{tabular}
\figcaption{Results of our phase-resolved spectral analysis.
Blackbody luminosity ($L_{\rm BB}$) is in units of $10^{35}\ \rm erg\ s^{-1}$,
and power-law flux ($F_{\rm PL}$) is in units of
$10^{-11}\ \rm erg\ s^{-1}\ cm^{-2}$ in the 3--79~keV band.
Gamma$_{\rm h}$ is the power-law index of the hard power-law component and
Gamma$_{\rm s}$ is for the soft component.
\label{fig:phspec}
}
\vspace{0mm}
\end{figure*}

We conducted a phase-resolved spectral analysis for
ten phase intervals to study distinct features in the pulse
profiles (see Figure~\ref{fig:pulseprofile} for pulse profiles).
We did not use the {\it Swift} XRT or {\it XMM-Newton} MOS data in this study
because their temporal resolutions were insufficient. The
{\it Chandra} and {\it XMM-Newton} PN data were phase-aligned
with the {\it NuSTAR} data by correlating the pulse
profiles.

We binned the {\it NuSTAR} and soft-band instrument
spectra to have at least 20 counts per spectral
bin, and froze the cross-normalization factors to those
obtained with the phase-averaged spectral fit. We fit the
spectra with the two models that we used for fitting the
phase-averaged spectrum: an absorbed blackbody plus
broken power law and an absorbed blackbody plus double
power-law model. We find that both models explain
the data well, having $\chi^2$/dof$\lapp$1.003 for dofs of 1413--1966.
The spectra vary with spin phase, having harder power-law spectra when
the flux is high. However, the detailed variation depends
on the spectral model used. We show the results in Figure~\ref{fig:phspec}.

\begin{figure*}
\vspace{-5 mm}
\centering
\begin{tabular}{cc}
\vspace{-5.0 mm}
\includegraphics[width=3.5 in]{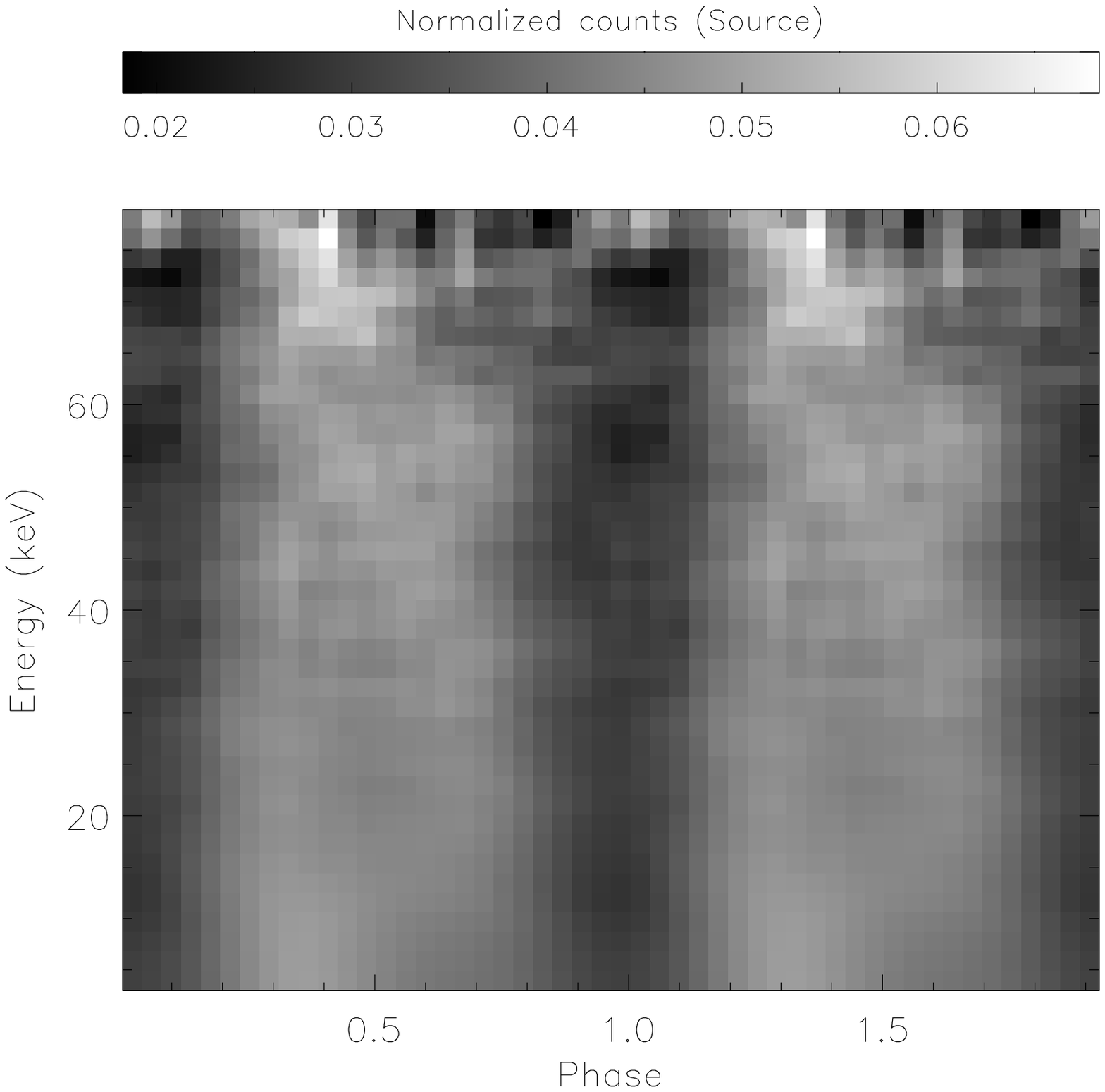} &
\includegraphics[width=3.5 in]{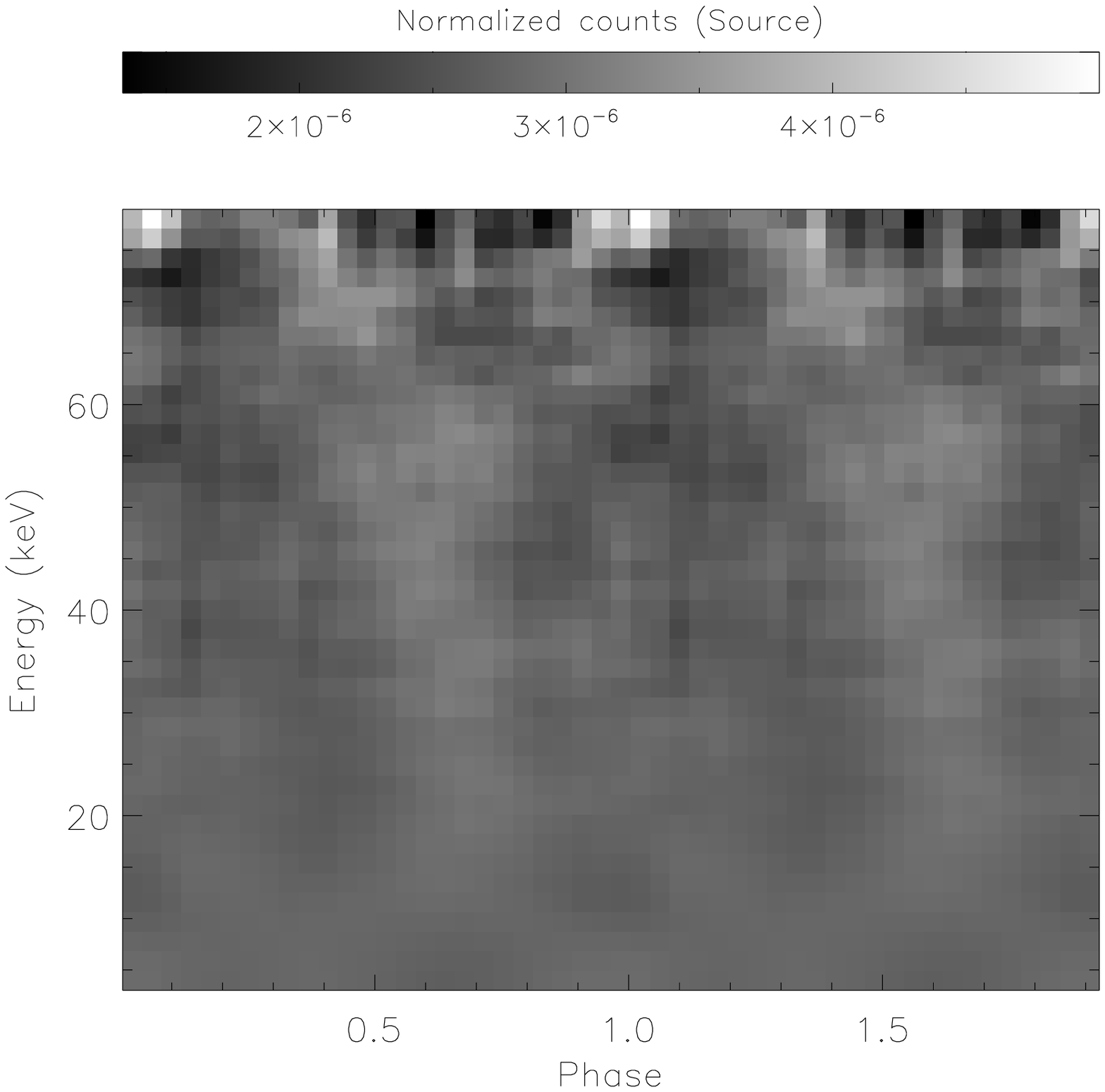}
\end{tabular}
\vspace{-5.0 mm}
\figcaption{Background-subtracted energy-phase count images in the
3--79~keV band produced using 25 phase bins and 40 energy bins.
Counts in each pixel were divided by the phase-integrated counts
in the same energy bin (left), and then by the energy-integrated counts
in the same phase bin (right). Two phases are displayed for clarity.
\label{fig:phaseenergy}
}
\end{figure*}

In order to see if there is a spectral feature that shifts
with energy as was seen in SGR 0418+5729 \citep[][]{tem+13},
we produced an energy-phase image using 25 phase
bins and 40 energy bins in the 3--79~keV band. We first
divided counts in each pixel in the energy-phase 2-D map
by the phase-integrated counts in the same energy bin
and present it in the left panel of Figure~\ref{fig:phaseenergy},
which shows similar structures to the energy-resolved pulse profiles
(Figure~\ref{fig:pulseprofile}). We then divided the map further by the
energy-integrated counts in the same phase bin (Figure~\ref{fig:pulseprofile})
in order to have a better contrast, and find no clear phase-dependent
feature in the image (Figure~\ref{fig:pulseprofile} right). We also
tried different binning and found the same results.

We measured the pulsed spectrum in the 0.5--79~keV
band in order to see if it is significantly harder than
the phase-averaged spectrum as seen in other hard X-ray
bright magnetars \citep{khh+06}. We grouped
the spectra to have at least 200 counts per spectral bin
and subtracted the spectrum in the phase interval 0.9--1.1
(the DC level) from the phase-averaged spectra obtained
in Section~\ref{sec:phavspec}. We then jointly fit the broad-band
spectrum with a power-law model letting the cross normalization
constants vary. We used {\tt lstat} and $\chi^2$
statistics and found that they give consistent results.

A simple power-law model with a photon index of 1.8
fits the 0.5--79~keV data well (reduced $\chi^2$/dof$<$1).
In order to verify the measurements for the pulsed spectrum
above 15~keV \citep[$\Gamma = 0.72 \pm 0.15$,][]{khh+06}, we
restricted the fit range to high energies. As the lower energy
cutoff is increased, the power-law index decreases,
consistent with spectral hardening. The results are summarized
in Table~\ref{ta:spec}.

We also estimated pulsed fractions in the hard band using
the spectra (defined as the ratio of pulsed and total
flux densities) in order to compare with those in Figure~\ref{fig:pulsedFrac}.
We fit the total and the pulsed spectra ($\gapp$15~keV) to
single power-law models. The power-law index of the total
spectrum above 15~keV is $1.19 \pm 0.02$ ($1.13 \pm 0.02$ above
20~keV), slightly smaller than what we obtained using the
absorbed blackbody plus broken power-law model (see Table~\ref{ta:spec}).
The power-law index of the pulsed spectrum
above 15~keV is $1.28 \pm 0.12$ ($1.0 \pm 0.2$ above 20~keV), similar
to that of the total spectrum. This suggests that the
pulsed flux does not rapidly increase in the hard band,
as also seen in Figure~\ref{fig:pulsedFrac}.

\subsection{Spectral fits with the $e^{\pm}$ outflow model}
\label{sec:bmodel}
A13 found that the properties of the persistent
hard X-ray emission of 1E~1841$-$045 was consistent
with the coronal outflow model proposed by \citet{b13}.
The model envisions an outflow of relativistic
electron-positron pairs created by electric discharge near
the neutron star. The outflow fills the active ``$j$-bundle''
--- a bundle of closed magnetospheric field lines that carry
electric current \citep{b09}. The pair plasma
flows out along the magnetic field lines and gradually
decelerates as it scatters the thermal X-rays. It radiates
most of its kinetic energy in hard X-rays before the $e^{\pm}$
pairs reach the top of the magnetic loop and annihilate.
The magnetic dipole moment of 1E~1841$-$045 is estimated
from its spin-down rate, $\mu \approx 7 \times 10^{32}\rm \ G\ cm^3$,
assuming the neutron star radius to be 10 km. Similar to
A13 and \citet{hbd14}, we assume a simple geometry where the
$j$-bundle is axisymmetric around the magnetic dipole axis. However,
instead of assuming that the $j$-bundle emerges from
a polar cap, its footprint is allowed to have a ring shape.
This possibility was introduced in \citet{vhk+14}, because
the {\it NuSTAR} data for 1E~2259$+$586 favored a ring
footprint over a polar cap. The model has the following
parameters:
(1) the power $\lj$ of the $e^{\pm}$ outflow along the $j$-bundle,
(2) the angle $\alphamag$ between the rotation axis and the magnetic axis,
(3) the angle $\betaobs$ between the rotation axis and the observer's line of sight,
(4) the angular position $\thetaj$ of the $j$-bundle footprint, and
(5) the angular width $\Delta \thetaj$ of the $j$-bundle footprint.
In addition, the reference point of the rotational phase, $\phaseref$,
is a free parameter, since we fit the phase-resolved spectra.

We follow the method presented in \citet{hbd14}
and explore the whole parameter space by fitting the phase-averaged
spectrum of the total emission (pulsed+unpulsed) and phase-resolved
spectra of the pulsed emission. We use five
equally spaced phase intervals. {\it NuSTAR} data are fitted
above 10~keV, where the coronal outflow has to account
for most of the observed emission.

\begin{figure}[t]
\begin{center}
\begin{tabular}{cc}
\vspace{2.0 mm}
\ \ \ \ $p$-value & \\
\vspace{-4.0 mm}
\includegraphics[trim = 7.6cm 2.5cm 8.cm 0.6cm, clip, height=0.35\textwidth]{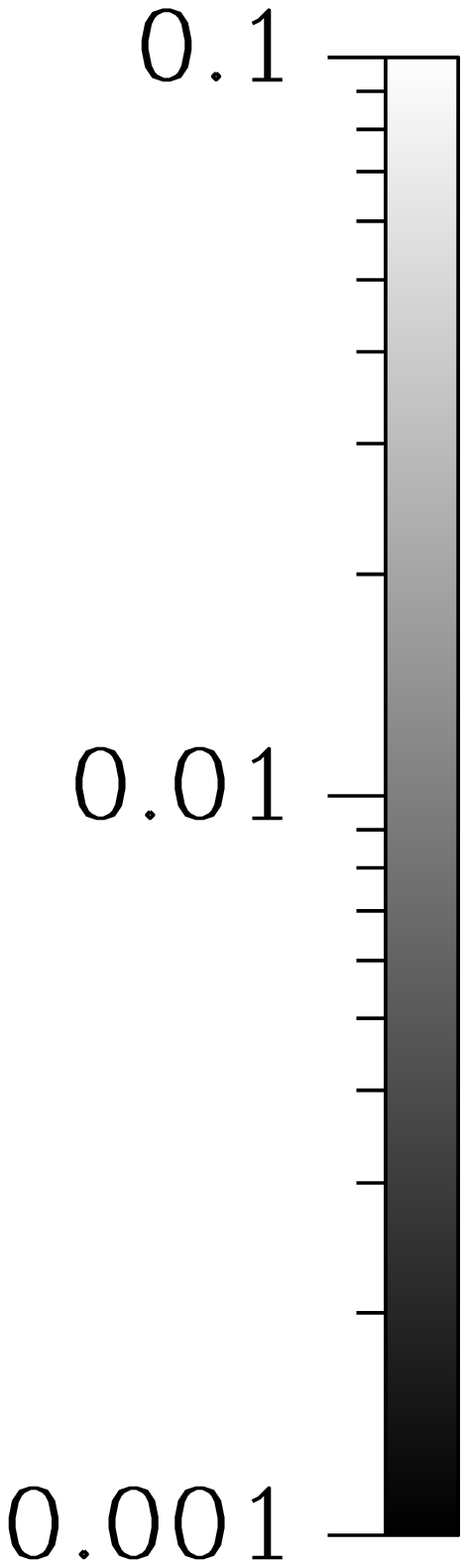}  & \hspace{-0.2cm}\includegraphics[trim = 0.cm 0.2cm 0.cm 0.7cm, clip, height=0.35\textwidth]{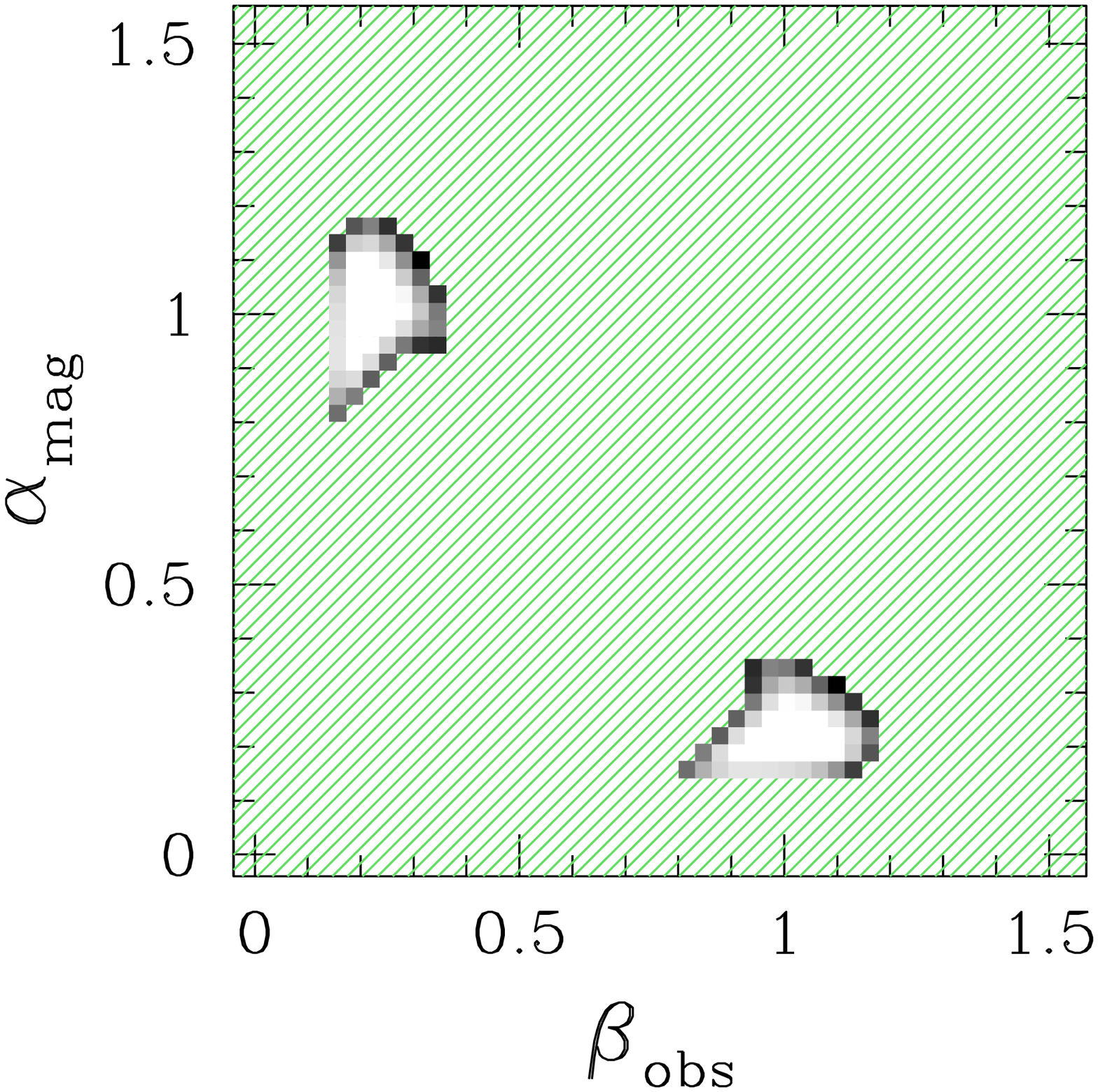}
\end{tabular}
\end{center}
\caption{
Map of $p$-values for the fit of the hard X-ray component with the
coronal outflow model; the $p$-values are shown in the plane of
 ($\alpha_{\rm mag}$, $\beta_{\rm obs}$) and maximized over the other parameters.
The $p$-value scale is shown on the left.
The hatched green region has $p$-values smaller than $0.001$;
the white region has $p$-values greater than $0.1$.
Interchanging the values of $\alphamag$ and $\betaobs$ does not change the
model spectrum, as long as the $j$-bundle is assumed to be axisymmetric.
Therefore, the map of $p$-values is symmetric about the line of $\betaobs=\alphamag$.
}
\label{fig_pvalue}
\end{figure}

Figure~\ref{fig_pvalue} shows the map of the $p$-value of the fit in the
$\alphamag$-$\betaobs$ plane. The acceptable model is clearly
identified in this map.\footnote{There are in fact two
solutions because interchanging the values of $\alphamag$ and $\betaobs$ does not
change the model spectrum, as long as the $j$-bundle is assumed to be axisymmetric.}
It has $\alphamag = 0.25 \pm 0.15, \betaobs = 1.0 \pm 0.2$,
$\thetaj = 0.24 \pm 0.02$, and $\Delta \thetaj/\thetaj > 0.26$,
consistent with a polar cap. The corresponding magnetic flux in the $j$-bundle is
$(2.5 \pm 0.4) \times 10^{26}\rm \ G\ cm^2$. The power
dissipated in the $j$-bundle is $\lj = (6 \pm 1) \times 10^{36}\rm \ erg\ s^{-1}$.
Most of the released energy is radiated in the MeV band
(peaking at $\sim$6\,MeV) and is not seen to {\it NuSTAR}.
Using the obtained best-fit model for the hard X-ray
component, we have investigated the remaining soft X-ray
component. The procedure is similar to that in A13
and \citet{hbd14}; we freeze the best-fit parameters of the outflow
model and fit the spectrum in the 0.5--79~keV band using
the {\it NuSTAR}, {\it Swift}, {\it Chandra}, and {\it XMM-Newton} data.
As in A13, we find that the 0.5--79~keV spectrum
is well fitted by the sum of two blackbodies (which
dominate below 10~keV) and the coronal outflow emission
(which dominates above 10~keV). The cold and hot blackbodies
have luminosities $L_c = 2.2 \pm 0.1 \times 10^{35}\rm \ erg\ s^{-1}$,
$L_h = 9.8 \pm 1.3 \times 10^{34}\rm \ erg\ s^{-1}$ and temperatures
$kT_c = 0.45 \pm 0.01\rm  \ keV$, $kT_h = 0.75 \pm 0.02\rm \ keV$.
Note that these values are different from those we obtained with the
phenomenological models in Table~\ref{ta:spec}.
\medskip

\section{Swift monitoring observations}
\label{sec:swiftobs}
We report below on {\it Swift} monitoring observations for
spectral and temporal behavior of the source on long
timescales. The observations were taken with the X-Ray
Telescope (XRT) using Windowed-Timing (WT) mode
for all observations, which have been conducted once every
two to three weeks since 2011 July, except when the
source was in Sun-constraint from mid-November to mid-February
each year. In total, 68 observations (not listed
in Table 1) having $\sim$266 ks of summed exposure were analyzed.
The {\it Swift} data were processed with {\tt xrtpipeline}
using the HEASARC remote
CALDB.\footnote{http://heasarc.nasa.gov/docs/heasarc/caldb/caldb\_remote\_acce\
ss.html}

\subsection{Data Analysis and Results for the Swift Monitoring Observations}
\label{sec:swiftmonitoring}

\begin{figure}
\vspace{-2.5 mm}
\centering
\includegraphics[width=3.6 in]{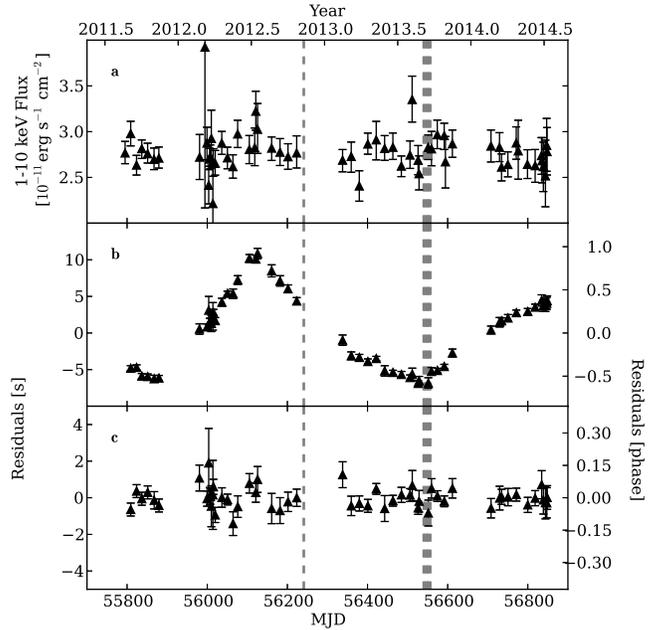} 
\vspace{-5 mm}
\figcaption{Results of the {\it Swift} monitoring campaign for flux and
timing. (a) 1--10~keV flux, (b) timing residuals after fitting out
$\nu$ and $\dot \nu$,
and (c) after fitting out twelve frequency derivatives.
Vertical dashed lines show the periods when {\it NuSTAR} observations were taken.
\label{fig:swift}
}
\end{figure}

To investigate the spectrum of 1E~1841$-$045 in the
monitoring observations, we extract spectra for each
observation using a 10-pixel (24$''$) long strip centered on
the source. An annulus of inner radius 75-pixel, and outer
radius 125-pixel centered on the source was used to extract
background spectra. The spectra were
grouped to have a minimum of 20 counts per bin. The
spectra were fitted with a photoelectrically absorbed
blackbody with an added power-law component, using
the {\tt tbabs(bbody+pow)} model in {\tt XSPEC} 12.8.1, with $N_{\rm H}$
held fixed at $2.05 \times 10^{22}\rm \ cm^{−2}$, the value we obtained in
Section~\ref{sec:phavspec}. No significant change in source 1--10~keV
flux was observed over the monitoring period of $\sim$3 years
($\chi^2$/dof=49/59), including the {\it NuSTAR}-observed bursting period.
The result is shown in Figure~\ref{fig:swift} (a).

We also searched all the {\it Swift} observations for bursts
by binning the source region light curves above 1~keV into
0.01~s, 0.1~s, and 1.0~s bins. The counts in each bin were
compared to the mean count rate of its GTI,
assuming the Poisson distribution. We found
no significant bursts in the {\it Swift XRT data}. Note
that the {\it Swift} observations did not cover the
{\it NuSTAR}-detected bursts times presented in Table~\ref{ta:bursts}.

In order to derive the timing solution and to search for
glitching activity, we barycentered the source events using
the location of 1E~1841$-$045, R.A.=$18^h41^m19.34^s$,
Decl.=$-4^\circ  56' 11\farcs2$.
We then extracted Time-Of-Arrivals (TOA) using a
Maximum Likelihood (ML) method \citep[see][]{lrc+09, snl+12}.
The ML method compares
a continuous model of the pulse profile to the profile
obtained by folding a single observation. These TOAs were
fitted to a pulse arrival model (e.g., the quadratic function
in Section~\ref{sec:timing}) using the TEMPO2 \citep{hem06} 
pulsar timing software package.
We find that a timing model consisting of $\nu$ and $\dot \nu$ does
not fit the data well (Figure~\ref{fig:swift} (b) and
``Fit~1'' in Table~\ref{ta:swiftsol})
as was already observed in the previous
{\it RXTE} monitoring observations \citep[e.g., see][]{dk14}.
We need to include twelve frequency derivatives
in order to achieve an acceptable fit (i.e., $\chi^2$/dof$\sim$1)
with Gaussian residuals (Figure~\ref{fig:swift} (c) and
``Fit~2'' in Table~\ref{ta:swiftsol}).
Note that the timing solutions presented in Table~\ref{ta:swiftsol}
are valid only over the time interval of the monitoring campaign.

Motivated by the apparent `kink' in the residuals of the simple spin-down model
around MJD~56100, we attempted to fit a glitch at the epoch of the kink.
However, the data are better fit using a model with four frequency
derivatives (rms residual of 0.97~s) versus
a glitch model (rms residual of 0.99 s).
Therefore we do not need to invoke a sudden event to explain the measured TOAs.
We present our best timing solutions in Table~\ref{ta:swiftsol}.

\begin{table}
\centering
\centering
\caption{Timing Parameters for 1E$\;$1841$-$045.}
\label{ta:swiftsol}
\begin{tabular}{ c c }
\hline
\hline
\rule{0pt}{3ex}
RA (J2000)              & $18^h41^m19.34^s$\\
\rule{0pt}{3ex}
DEC (J2000)            & $-4^\circ  56' 11.2''$\\
\rule{0pt}{3ex}
MJD Range               & 55795-56799  \\
\rule{0pt}{3ex}
Epoch (MJD)             & 56300\\
\hline
\multicolumn{2}{c}{Fit 1} \\
\hline
\rule{0pt}{3ex}

$\nu$ (s$^{-1}$) & 0.084 806 860 6(9) \\
\rule{0pt}{3ex}
$\frac{d^1\nu}{dt^1}$ (s$^{-2}$) & $-$2.9121(8)$\times 10^{-13}$ \\
\rule{0pt}{3ex}
RMS (s) &4.82 \\
$\chi^2$/dof &  2889.05/53 \\
\hline
\multicolumn{2}{c}{Fit 2} \\
\hline
\rule{0pt}{3ex}

$\nu$ (s$^{-1}$) &0.084 806 897(4))  \\
\rule{0pt}{3ex}
$\frac{d^1\nu}{dt^1}$ (s$^{-2}$) & $-$2.985(12)$\times 10^{-13}$ \\
\rule{0pt}{3ex}
$\frac{d^2\nu}{dt^2}$ (s$^{-3}$) & 6.8(23)$\times 10^{-22}$ \\
\rule{0pt}{3ex}
$\frac{d^3\nu}{dt^3}$ (s$^{-4}$) & 4.6(8)$\times 10^{-28}$ \\
\rule{0pt}{3ex}
$\frac{d^4\nu}{dt^4}$ (s$^{-5}$) & $-$8.7(16)$\times 10^{-35}$ \\
\rule{0pt}{3ex}
$\frac{d^5\nu}{dt^5}$ (s$^{-6}$) & $-$2.5(5)$\times 10^{-41}$ \\
\rule{0pt}{3ex}
$\frac{d^6\nu}{dt^6}$ (s$^{-7}$) & 6.0(12)$\times 10^{-48}$ \\
\rule{0pt}{3ex}
$\frac{d^7\nu}{dt^7}$ (s$^{-8}$) & 1.2(3)$\times 10^{-54}$ \\
\rule{0pt}{3ex}
$\frac{d^8\nu}{dt^8}$ (s$^{-9}$) & $-$3.4(7)$\times 10^{-61}$ \\
\rule{0pt}{3ex}
$\frac{d^9\nu}{dt^9}$ (s$^{-10}$) & $-$4.5(10)$\times 10^{-68}$ \\
\rule{0pt}{3ex}
$\frac{d^{10}\nu}{dt^{10}}$ (s$^{-11}$) & 1.4(3)$\times 10^{-74}$ \\
\rule{0pt}{3ex}
$\frac{d^{11}\nu}{dt^{11}}$ (s$^{-12}$) & 9.5(22)$\times 10^{-82}$ \\
\rule{0pt}{3ex}
$\frac{d^{12}\nu}{dt^{12}}$ (s$^{-13}$) & $-$3.3(8)$\times 10^{-88}$ \\

\rule{0pt}{3ex}
RMS (s) &0.43 \\
$\chi^2$/dof & 49.34/42 \\
\hline
\hspace{0.5 mm}
\end{tabular}\\
\footnotesize{{\bf Notes.} All errors are TEMPO2 reported 1$\sigma$ errors.
}\\
\end{table}
\medskip
\section{Discussion}
\label{sec:disc}
The new 350-ks observation of 1E 1841−045 by {\it NuSTAR}
allowed a significantly better study of its
persistent emission and the serendipitous detection of X-ray
bursts. Below we discuss the results and compare them
with observations of other magnetars.

\subsection{The X-ray Bursts and the Tails}
Magnetars often show bursting behavior in the X-ray
band, which may be caused by instabilities inside the
neutron star or its magnetosphere \citep[][]{td95,l03,wt06}.
The time profiles and spectra of bursts show significant
diversity. \citet{wkg+05} suggested that there are
two types of magnetar bursts, one having significant
tail emission and the other having orders of magnitude
smaller tail emission. The authors attributed the former
to crustal activity and the latter to magnetospheric
activity. \citet{lwg+03} found a strong correlation between
bursts and tail energies in the magnetar SGR~1900$+$14.

The X-ray bursts from 1E~1841$-$045 in 2013 September
have very short rise and fall times, with $T_{\rm 90}$ of
0.01--0.6 s, and hard spectra ($\Gamma$=1--2 or $kT\sim$3--5~keV).
The blackbody temperature we measured
for the burst 5 ($4.8\pm0.5$\,keV; see Table~\ref{ta:burst90})
is consistent with that of the colder blackbody measured with the
GBM \citep[$kT_l = 5.3\pm0.2$~keV;][]{cxc14}.

\citet{ks10} reported detection of emission lines at 27~keV,
40~keV and 60~keV in the 2010 May
burst spectrum with {\it Swift} BAT, although these are argued
against later by \citet{lkg+11}. Interestingly, all
the lines are in the {\it NuSTAR} band, and could be detected
by {\it NuSTAR} if they appeared again.
However, we do not see evidence of line emission. Therefore,
we estimate 90\% upper limits on any 27\,keV Gaussian line flux
to be 0.24~photons~$\rm cm^{-2}\ s^{-1}$ and
0.11~photons~$\rm cm^{-2}\ s^{-1}$ in the brightest burst spectrum
for the blackbody and the power-law
continuum models, respectively. Note that we are not able to
compare our results with those of \citet{ks10} since they did
not present the line flux.

An extended tail is reliably detected only in energetic
burst 5, and we find a hint of a tail in burst 6. Note that
these two bursts were also observed by the GBM,
and had significant flux
above the {\it NuSTAR} band \citep[][]{cxc14}; they are the two most
energetic bursts in our sample. Thus, the tail brightness
and the burst energy we measure for the 1E~1841$-$045 bursts
seem to agree qualitatively with the correlation reported
for SGR~1900$+$14 and SGR~1806$-$20 \citep{lwg+03, gwk+11}.
A similar trend was also seen in the recent bursts
from 1E~1048.1$-$5937 \citep{akb+14}.
We note however that the large energy
seen by GBM in bursts 5 and 6 indicates a tail-to-burst
ratio ($E_{\rm tail}/E_{\rm burst}\sim2\times10^{-2}$) that is much lower than
in 1E~1048.1$-$5937 ($E_{\rm tail}/E_{\rm burst}\sim5$--$60$).
This confirms the known diversity of tails of magnetar bursts
\citep[][]{kgw+04, wkg+05}. Whether the tail-to-burst ratio
follows a bimodal or a random distribution
is not yet clear, and further investigation is required.

In contrast to the burst tails observed by {\it NuSTAR} for
1E~1048.1$-$5937, the burst tail in 1E~1841$-$045 shows
no clear correlation between spectral hardness and flux.
This correlation was also absent in some of the long-term
(months to years) flux relaxation of other magnetars
\citep[e.g.,][]{akt+12}.

The flux evolution in the tail of burst 5 followed a
power-law decay with a decay index of $0.45\pm0.10$. The flux
decay is similar to the tails of bursts from SGR~1900$+$14
\citep[decay index of 0.43--0.7;][]{lwg+03}. A significantly
faster decay was observed for burst tails in
1E~1048.1$-$5937 \citep[decay index of 0.8--1;][]{akb+14}.
It is possible that the tail contains many unresolved
weaker bursts which affect the observed flux decay. It is
still unclear what controls the resulting decay rate and
why it is significantly different in 1E~1048.1$-$5937 and
1E~1841$-$045.

Furthermore, we find that the tail spectra in
1E~1841$-$045 are similar to (or slightly softer than)
the persistent emission. In contrast, the tail spectra in
1E~1048.1$-$5937 were significantly harder than its persistent
emission. This further contributes to the diversity of
magnetar bursts. For instance, 1E~2259$+$586 exhibited
bursts with and without ks-long tails, and the tail emission
was observed to soften with decreasing flux \citep[][]{kgw+04}.
The observed diversity of magnetar bursts is not well explained
by current theoretical models.

{\subsection{Pulse Profile and Pulsed Fraction}
\label{sec:timingdisc}
It is known that the pulse profiles of magnetars can
look significantly different in different energy bands \citep[][]{dkh+08}.
This is also true for 1E~1841$-$045 (Figure~\ref{fig:pulseprofile}).
An interesting feature of the pulse profile is
the double-peaked structure in the narrow band of $\sim$24--35~keV.
A13 found this in the previous {\it NuSTAR} observation,
and suggested that it may be
caused by a spectral feature. The new long observation
confirms the double peaked structure. It shows however
that the change in the pulse profile is not as sharp as was
suggested previously, and its shape may be more complicated,
as a hint of another peak appearing between the
two peaks is seen in several energy bands (e.g., in the
33--35~keV profile in Figure~\ref{fig:pulseprofile}).

The pulsed fraction (and the rms pulse amplitude) increases
with photon energy below $\sim$8~keV (Figure~\ref{fig:pulseprofile}).
We found no significant increase in the pulsed fraction
above 8~keV, in contrast with previous measurements
\citep{khh+06}. This conclusion is insensitive to
the choice of energy bins in Figure~\ref{fig:pulseprofile},
which can affect the measured values in individual bins,
but not the general trend. Our spectral analysis also suggests that the
pulsed and steady components of the flux have similar
power-law photon indices at high energies $>$15~keV
(see Section~\ref{sec:phresspec} and Table~\ref{ta:spec}).
In the coronal outflow model,
this implies that one of the angles, $\alphamag$ and $\betaobs$,
is small, or the emission region ($\thetaj$) is broad.

\subsection{Spectra}
\label{sec:ttspecdisc}
We have measured the phase-averaged, phase-resolved
and pulsed spectra of the source. Our best-fit parameters
are slightly different from those reported by A13.
The use of the new CALDB for
the {\it NuSTAR} data may have some effect on the obtained
spectral shape, however we showed in Section~\ref{sec:phavspec}
that this effect is small.

We do not see any significant change in the source flux
among the observations taken over one year despite the fact
that bursts were detected in some observations but not in the others.
We further compared the {\em NuSTAR}-measured spectrum
with that reported previously \citep[$\Gamma=1.32\pm0.11$ and
$L_{\rm 10-100 keV}=3.0\times 10^{35}\rm \ erg\ s^{-1}$
for an assumed distance of 6.7 kpc;][]{khh+06}, and find that
our measurement ($\Gamma=1.37\pm0.01$ and
$L_{\rm 10-100 keV}=3.0\times 10^{35}\rm \ erg\ s^{-1}$)
is fully consistent with the previous values, suggesting that the
source hard-band spectrum has been stable over 10 years.
The same trend in the soft band (1--10~keV) is seen
in the {\it Swift} monitoring data (see Figure~\ref{fig:swift}).
Stability in the soft-band pulsed flux has been reported by
\citet{dk14} and \citet{zk10} for a longer period.
We note that \citet{ks10} reported a possible increase
in the soft-band flux for 1E~1841$-$045
($<$10~keV) associated with the source burst activity.
However, \citet{lkg+11} showed that
there is no significant change in the soft-band flux over
a 1400-day interval, including the burst period reported
by \citet{ks10}.

The new measurements of the phase-resolved spectrum
agree with the previous {\it NuSTAR} observation. The spectrum
is harder near the pulse peaks, with smaller $\Gamma_{\rm h}$ and
greater $E_{\rm break}$. We also note a possible hardening of the
phase-averaged spectrum above $>$15~keV.
However, we cannot reliably measure the spectrum
curvature in the
hard X-ray band due to the statistical noise.

We have searched for the spectral feature that was suggested
as a possible explanation for the change in the
pulse profile near 30~keV (A13). We find that
the phase-resolved spectra are all well fit by a blackbody
plus broken power law or a blackbody plus double
power law; no emission or absorption line is required or
hinted at by the fit residuals. We also searched for a spectral
feature that shifts with rotational phase as was seen
in SGR~0418$+$5729 \citep{tem+13}, but did not find
such a feature (Figure~\ref{fig:phaseenergy}).

\subsection{Outflow model}
The more accurate phase-resolved spectrum obtained
from the 350-ks {\it NuSTAR} observation provides
a new opportunity to apply the coronal outflow model.
We found that the model
still fits the observed hard X-ray spectrum, and it does
so in a small region of the parameter space (Figure~\ref{fig_pvalue}),
which allowed us to estimate the size of the active $j$-bundle,
the dissipated power in the magnetosphere,
and the angles between the magnetic axis, the rotation axis,
and the line of sight. The increase in exposure by a factor
$\sim$6 excluded at more than 3$\sigma$ level the second solution
for the outflow model that was found in A13.

Interestingly, adding the new free parameter $\Delta \thetaj$ does
not introduce new acceptable solutions (with a $p$-value
above $10^{-3}$) and does not significantly affect the best
solution --- the best fit shows that the footprint of the
$j$-bundle on the star is a broad ring, hardly distinguishable
from a polar cap. This contrasts with 1E~2259$+$586,
where a thin ring with $\Delta \thetaj/\thetaj < 0.2$
is clearly statistically favored over a polar cap. This diversity may point
to different distributions of the crustal magnetic stresses
which are responsible for the magnetospheric twisting.

The soft X-ray component (below $\sim$10~keV) is well
fitted by the sum of two blackbodies. The best-fit
model is similar to that found in A13. The
cold blackbody covers a large fraction of the star area,
$\mathcal{A}_{c} \approx 0.42 \mathcal{A}_{\rm NS}$,
where $\mathcal{A}_{\rm NS}$ is the area of the neutron
star with an assumed radius $R_{\rm NS} = 10\rm \ km$. The
emission area of the hot blackbody,
$\mathcal{A}_{h} \approx 0.024 \mathcal{A}_{\rm NS}$,
is comparable with the area of the outflow footprint
$\mathcal{A}_{j} = \pi \sin^2 \theta_j \simeq 0.014 \ (\thetaj / 0.24)^2$.
This is consistent
with the coronal outflow model, where the footprint of
the $j$-bundle is expected to form a hot spot, as some particles
accelerated in the $j$-bundle flow back to the neutron
star and bombard its surface.

Figure~\ref{fig_pvalue} suggests that the coronal outflow 
correctly describes the hard X-ray source as a 
decelerating $e^\pm$ outflow ejected from a discharge zone near the star. 
The outflow parameters inferred from the fit of the phase-resolved data have rather 
small statistical uncertainties, and the results reveal
a puzzling feature of 1E~1841$-$045.
Using Equation~(48) in \citet{b09}, one can infer
the discharge voltage in the active $j$-bundle:
$\Phi \approx 10^{11} \psi^{-1} \ \mathrm{V}$ where $\Psi$
is the twist implanted in the
magnetosphere, which does not exceed $\psi_{\rm max}\sim 3$
\citep[][]{pbh13}. Due to our refined constraint on $\thetaj$ and
the strong dependence of $\Phi$ on $\thetaj$ ($\Phi \propto \thetaj^4$),
the inferred voltage is one order of magnitude higher than that given
in A13. The high voltage is surprising in two ways:
(1) it exceeds the expected threshold for $e^{\pm}$ discharge \citep{bt07}
by at least a factor of 10;
(2) it implies a short timescale for ohmic dissipation of
the magnetospheric twist $t_{\rm diss} \approx 0.1\, \psi^2$~yr
\citep[Equation (50) in][]{b09}. Thus, 
without continued energy supply, the  $j$-bundle is expected to 
untwist
on a year timescale or faster, which
is not observed --- the persistent X-ray emission from
1E~1841$-$045 has been stable for at least one decade.

This puzzle is related to a more general question: why are some magnetars
transient and others persistent? The magnetospheric activity should be fed by 
magnetic energy pumped from the star by its surface motions. 
The surface motions are caused by the crust yielding to 
accumulating magnetic stresses inside the magnetar. Energy supply to 
the twisted magnetosphere may or may not be intermittent, depending on the 
mechanism of the crustal motions.
\citet{bl14} have recently shown that the crust can yield
through a thermoplastic instability
which launches a slow wave resembling the deflagration
front in combustion physics. The thermoplastic
wave rotates the crust and twists the magnetosphere.
This mechanism naturally triggers the outbursts observed in transient magnetars,
yet it is still unclear how the quasi-steady state observed in 1E~1841$-$045 is formed.
It could be formed by frequently repeated energy supply to the $j$-bundle combined 
with some form of feedback on its untwisting rate.
\medskip

\subsection{Swift monitoring observations}
\label{sec:swiftdisc}
We did not detect any changes in the 1--10~keV source
flux during the three years of {\it Swift}-XRT monitoring since 2011.
In particular, the
flux did not change significantly during the bursting period,
which is consistent with the stability of persistent
emission in the {\it NuSTAR} data. Our measurements do
not agree with the observation that the source persistent emission
properties changed due to bursts \citep[][]{ks10}. Note that
\citet{lkg+11} also observed no change in the persistent properties
of the source after bursts.
The flux stability also implies that the
source seems to be in a perpetually bursting state, and
perhaps this is a difference between the classical bright
AXPs monitored with {\it RXTE} and `transient' sources, i.e.,
perhaps the bursting indicates that there is more
heat dissipating and hence more magnetic activity in this magnetar, consistent
with the higher $L_{\rm X}$ in ``quiescence''.

The pulse timing behavior of 1E~1841$-$045 is known
to be noisy, and \citet{dk14} had to use five
derivative terms to fit the timing data.
Our timing solution requires twelve frequency derivatives.
Note that although \citet{dk14} used a much longer data span (13 yr),
they break the data into smaller pieces which span 1--3 yr,
similar to ours.
The {\em Swift} monitoring data require more frequency derivatives
than the {\em RXTE} data do due to the large kink at MJD 56100 which
may indicate a glitch. We investigated the possibility
that the kink is caused by a glitching activity but did not find
any evidence of a glitch during the monitoring period of
3 years including the bursting period.

\section{Conclusions}
\label{sec:concl}
During {\it NuSTAR} observations of the magnetar
1E~1841$-$045, we detected six X-ray bursts from the
source. The bursts are short, $T_{\rm 90} < 1$~s, and bright.
A tail was observed after one burst which was most energetic
as measured with {\it Fermi} GBM. The tail emission
was similar to or softer than the persistent emission, and
the flux decay in the tail followed a power law with a decay
index of $\sim$0.5, with no clear spectral softening with
time. The properties of the tail emission are different
from those seen after recent {\it NuSTAR}-observed X-ray
bursts from the magnetar 1E~1048.1$-$5937 whose tails
decayed fast with flux decay indices of 0.8--0.9 and had
harder spectra than the persistent emission. The new observations
also yield detailed pulse profiles in different energy
bands, and show that the pulsed fraction does not increase
rapidly with energy, in contrast to previous observational
reports. We show that the source hard-band flux has been stable
over $\sim$10 years.
Using a {\it Swift} monitoring campaign,
we found that the 1--10~keV flux from the source has
been stable within $<$20\% and the source timing behavior
has been very noisy during a 3-year period,
in spite of the bursting behavior we have observed.

The X-ray spectra of 1E~1841$-$045 are well fitted by the coronal outflow
model of \citet{b13}. The new fit has provided
improved constraints on the angle between the magnetic
and rotation axes of the magnetar and the size of its
active $j$-bundle. Remarkably, the best fit implies fast
dissipation of the magnetic twist, in apparent contradiction
with the observed stability of X-ray emission, as
no long-term evolution has been detected in the persistent
emission of 1E~1841$-$045 for more than one decade
\citep[][A13]{khh+06}. This behavior
is distinct from the untwisting magnetospheres in
transient magnetars and presents a puzzle that is yet to be
resolved.

\medskip

This work was supported under NASA Contract No. NNG08FD60C,
and  made use of data from the {\it NuSTAR} mission,
a project led by  the California Institute of Technology, managed by the Jet Propulsion  Laboratory,
and funded by the National Aeronautics and Space  Administration. We thank the {\it NuSTAR} Operations,
Software and  Calibration teams for support with the execution and analysis of  these observations.
This research has made use of the {\it NuSTAR}  Data Analysis Software (NuSTARDAS) jointly developed by
the ASI  Science Data Center (ASDC, Italy) and the California Institute of  Technology (USA).
H.A. acknowledges supports provided by the NASA sponsored Fermi Contract NAS5-00147 and by
Kavli Institute for Particle Astrophysics and Cosmology.
V.M.K. acknowledges support
from an NSERC Discovery Grant and Accelerator Supplement, the FQRNT Centre de Recherche Astrophysique du Qu\'ebec,
an R. Howard Webster Foundation Fellowship from the Canadian Institute for Advanced
Research (CIFAR), the Canada Research Chairs Program and the Lorne Trottier Chair
in Astrophysics and Cosmology.
A.M.B. acknowledges the support by NASA grant NNX13AI34G.

\begin{appendix}

\section{COMPARISON BETWEEN ESTIMATORS OF PULSED FRACTION}
\label{sec:appendix1}
Some confusions on pulsed fraction of a pulsar have been arisen
mainly because different literatures use a different
estimator. In this appendix, we review several commonly used
estimators of pulsed fraction, and describe cautions to
be taken when using them. 

\begin{figure*}
\centering
\begin{tabular}{ccc}
\hspace{-7.0 mm}
\vspace{-5.0 mm}
\includegraphics[width=2.65 in, angle=0]{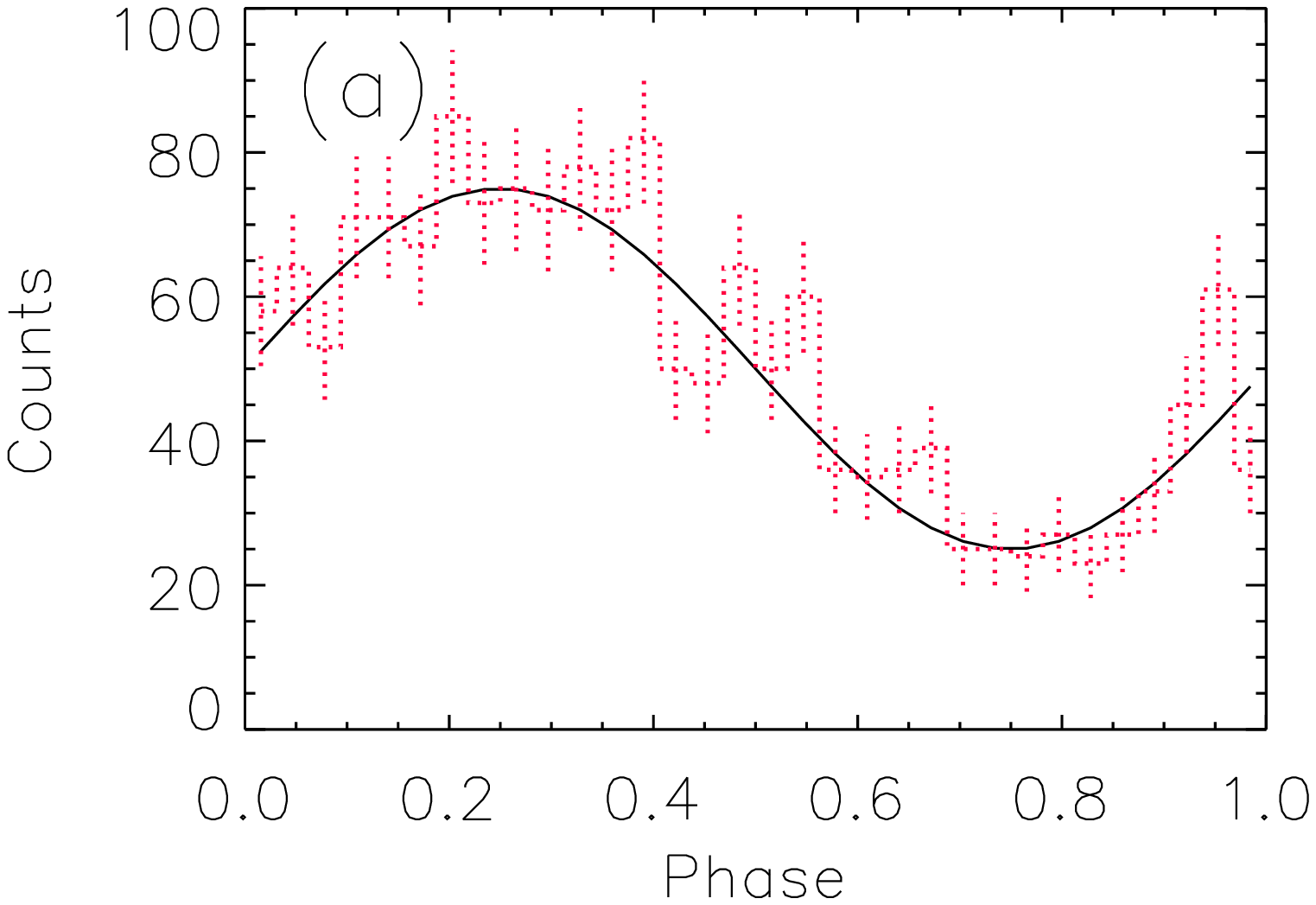} &
\hspace{-12.0 mm}
\includegraphics[width=2.65 in, angle=0]{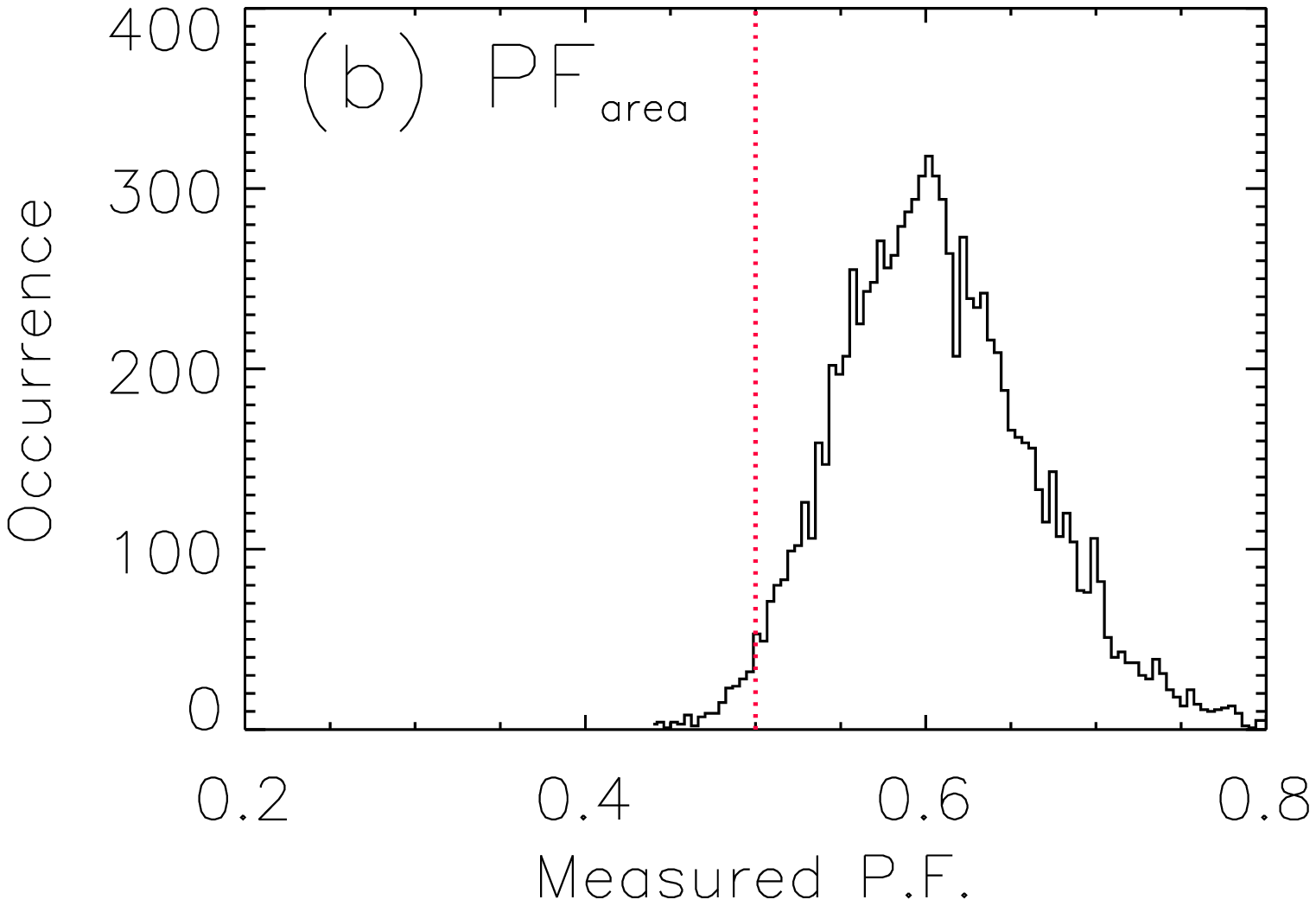} &
\hspace{-12.0 mm}
\includegraphics[width=2.65 in, angle=0]{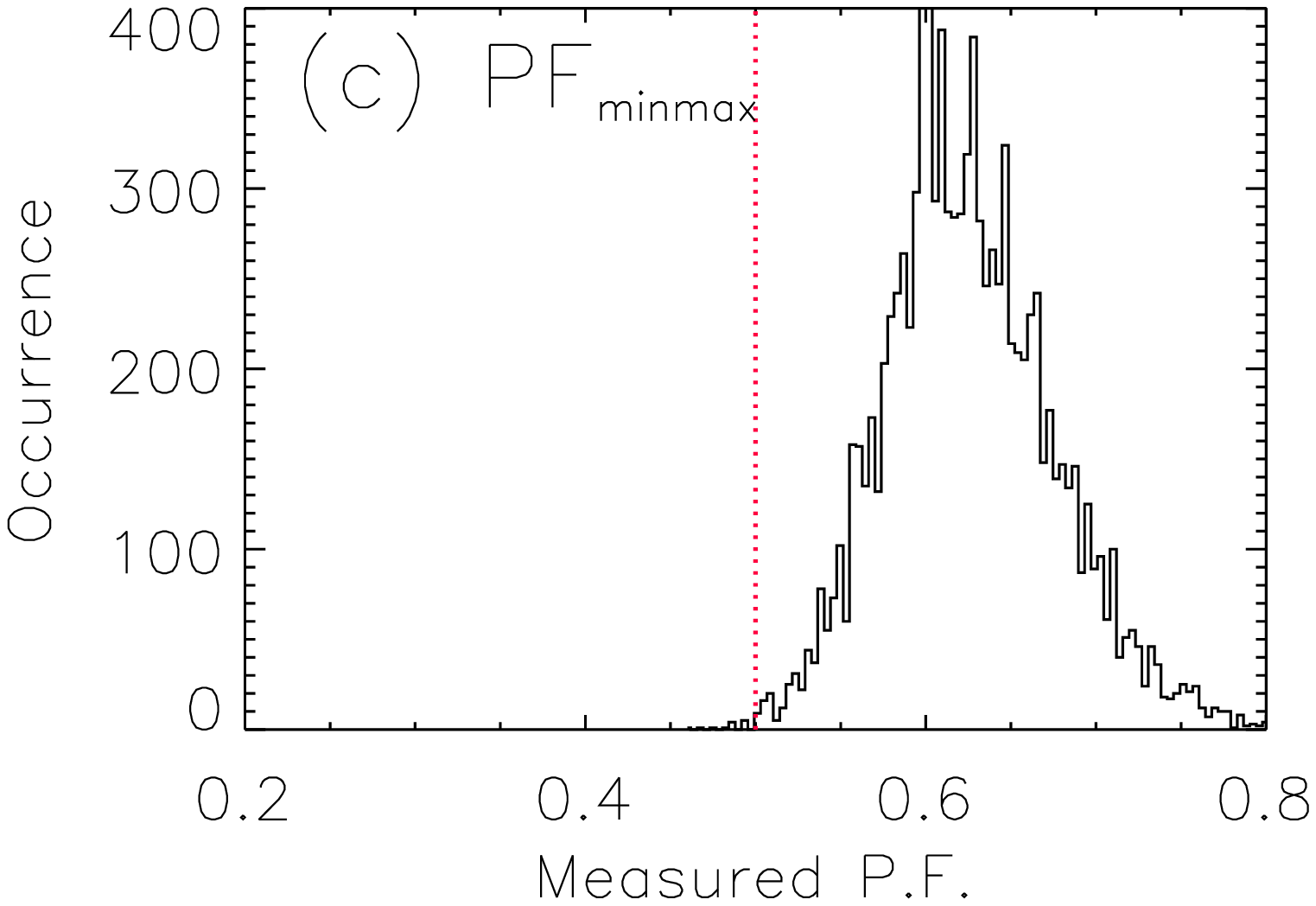} \\
\hspace{-7.0 mm}
\vspace{-3.0 mm}
\includegraphics[width=2.65 in, angle=0]{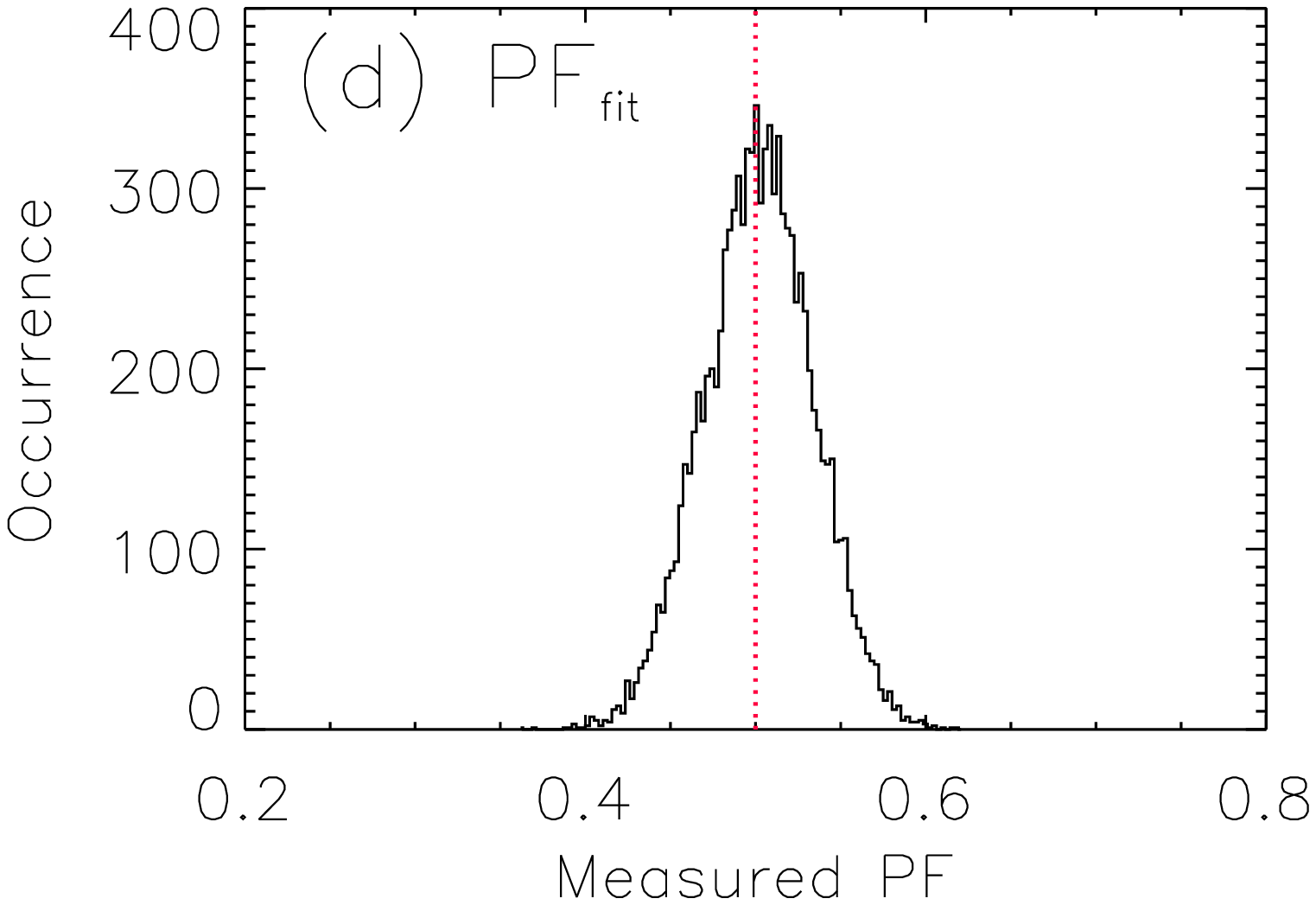} &
\hspace{-12.0 mm}
\includegraphics[width=2.65 in, angle=0]{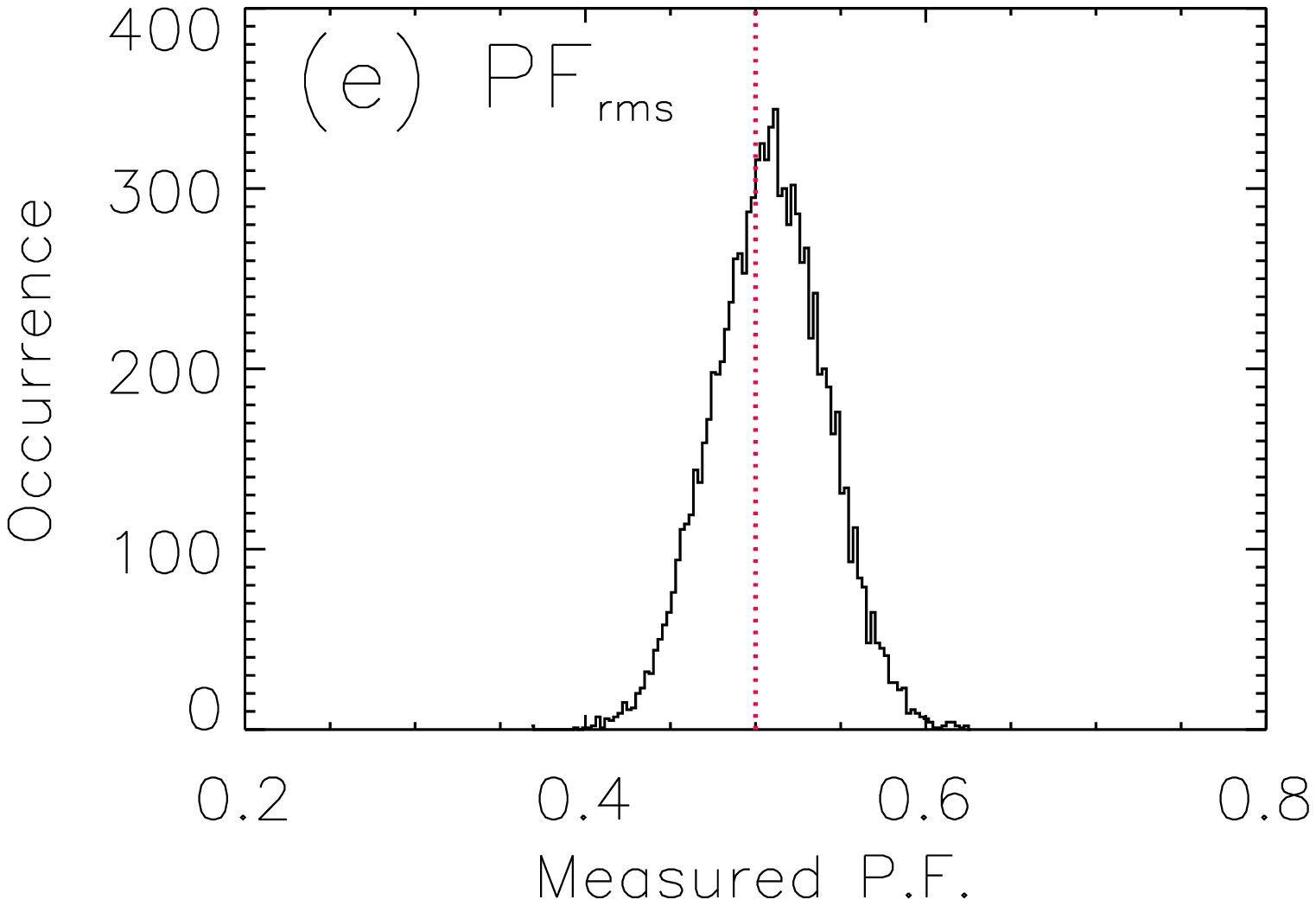} &
\hspace{-12.0 mm}
\includegraphics[width=2.65 in, angle=0]{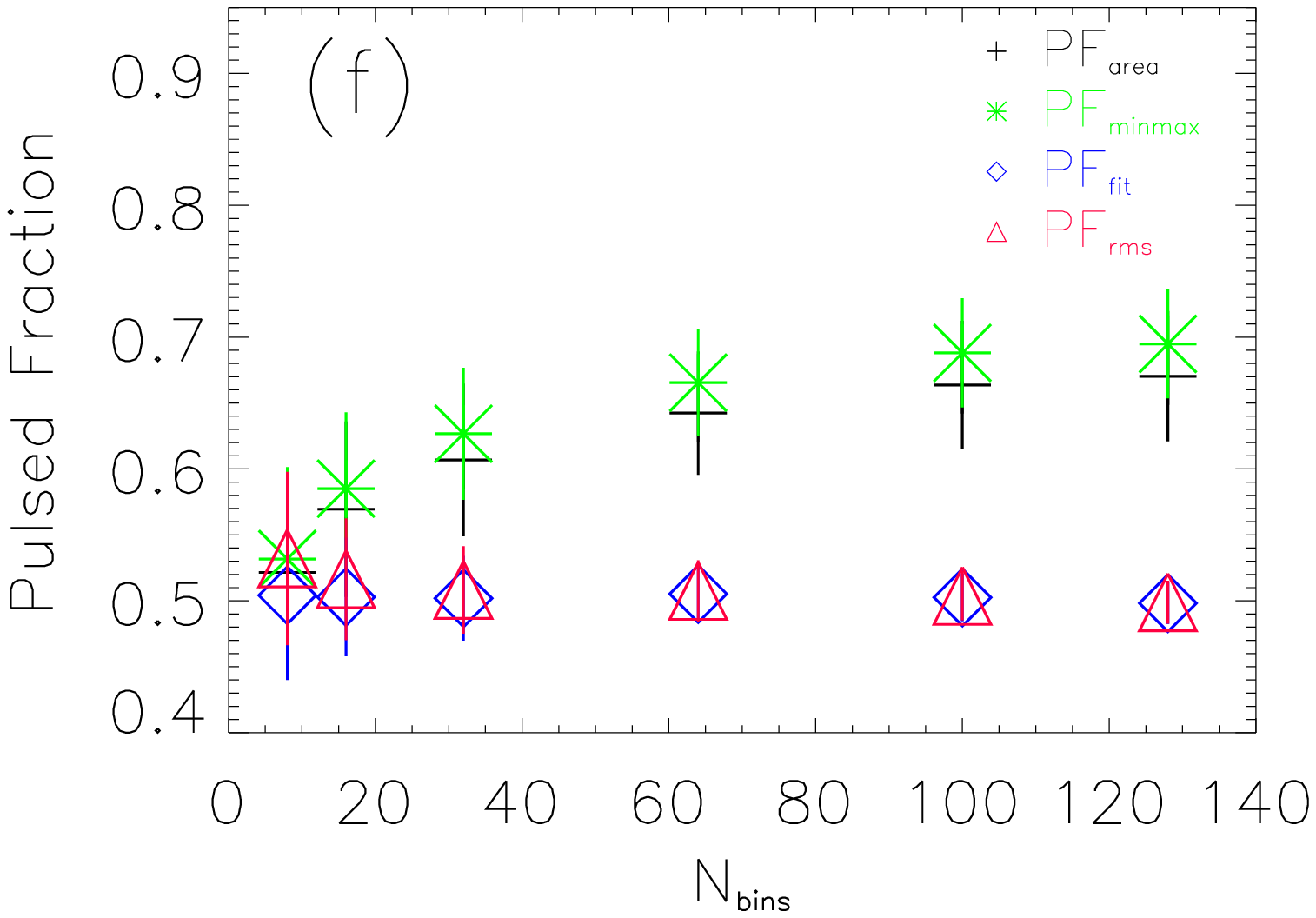} \\
\vspace{-3.0 mm}
\end{tabular}
\figcaption{Simulation results: (a) pulse profiles where the solid line
is a theoretical pulse profile, and the dotted curve shows a realization
of the theoretical pulse profile, and pulsed fractions measured using
(b) area fraction ($PF_{\rm area}$),
(c) min-max bin ($PF_{\rm minmax}$),
(d) a function fit ($PF_{\rm fit}$), 
(e) rms amplitude ($\sqrt{2}PF_{\rm rms}$), and
(f) for different binning.
Vertical red dotted line
in panels (b)-(f) shows the theoretical value for the pulsed fraction.
\label{fig:PFsim}
}
\vspace{0mm}
\end{figure*}

We consider four different estimators of pulsed fraction
commonly used in literatures: pulsed fraction measured by (1)
count area $PF_{\rm area}$ (Equation~\ref{eq:PFarea}),
(2) min-max counts,
$PF_{\rm minmax} = \frac{p_{\rm max}-p_{\rm min}}{p_{\rm max}+p_{\rm min}}$,
(3) fitting the light curve ($PF_{\rm fit}$), and
(4) calculating rms variation ($\sqrt{\frac{\sum_i(p_i - <p>)^2}{<p>^2}}$)
using truncated Fourier series and
subtracting Fourier power produced by noise in the pulse profile,
$\sigma_{a_k}^2 + \sigma_{b_k}^2$, from the Fourier amplitudes
\citep[$PF_{\rm rms}$, Equation~\ref{eq:PFrms}, see also][]{abp+14}.
As we show below, estimators (1) and (2) are biased upwards
since one has to select the minimum (and maximum)
count bin in the pulse profile. For example,
if the DC component of a pulse profile is relatively broad, one
of the DC bins will very likely fall below the ``true'' minimum
because of statistical fluctuation. Since one will use
that phase bin for $p_{\rm min}$, the resulting pulsed fraction will be
higher than the true value. Although this bias can be
avoided by holding the minimum phase ($\phi_{\rm min}$)
fixed and performing simulations, the spread in the measurements will
be larger in this case than in the case of picking up
the minimum bin as we show below.

\newcommand{\markap}{\tablenotemark{a}}
\begin{table}[t]
\vspace{-0.0in}
\begin{center}
\caption{Summary of measurements made with 10,000 simulations
\label{ta:PFsim}}
\vspace{-0.05in}
\scriptsize{
\begin{tabular}{ccccc} \hline\hline
Estimator	&Mean	&Spread ($1\sigma$)	& $\Delta_{\rm theory}^\markap$	& Notes	\\ \hline
$PF_{\rm area}$	& 0.61		& 0.06		& 0.11		& $\cdots$ \\ 
$PF_{\rm area}$	& 0.5		& 0.1		& 0.0		& Fixed $\phi_{\rm min}$   \\ 
$PF_{\rm minmax}$& 0.63		& 0.05		& 0.13		& $\cdots$   \\
$PF_{\rm minmax}$& 0.50		& 0.09		& 0.0		& Fixed $\phi_{\rm min}$   \\
$PF_{\rm fit}$	& 0.501		& 0.031		& 0.001		& $\cdots$   \\ 
$PF_{\rm rms}$	& 0.512		& 0.032		& 0.012		& $\cdots$  \\ 
$PF_{\rm rms}$	& 0.513		& 0.032		& 0.013		& No correction term  \\  \hline
\end{tabular}}
\end{center}
\vspace{-1.0mm}
$^{\rm a}${Difference between the theoretical value and the mean of the measurements.}\\
\vspace{-2.0 mm}
\end{table}

We conducted simulations to compare the estimators.
For the simulations, we used a sine plus constant function
($A\sin\phi + B$) for the theoretical pulse profile.
The pulsed fraction for this profile can be analytically calculated for all
the estimators, and is A/B for (1)--(3),
and A/(B$\sqrt{2}$) for (4). We carried out 10,000 realizations of the
pulse profile for $A$ = 25 and $B$ = 50 per phase bin having a total of
$\sim$800 events in the pulse profile (Figure~\ref{fig:PFsim} (a)),
and measured the pulsed fraction using the above estimators.

After each realization, we measured the pulsed fraction as defined above.
Note that for the estimator (3), we used
a sine plus a constant function for the fit.
We show the results in Figure~\ref{fig:PFsim} and Table~\ref{ta:PFsim}.
We find that estimators (1)
and (2) give a biased value with a chance of 98\% and 99.8\%,
respectively (see Figures~\ref{fig:PFsim} (b) and (c)). When
holding the minimum phase fixed at the theoretical minimum phase,
no significant bias is seen for both the estimators,
but the spread of the measurements increased (Table~\ref{ta:PFsim}).
The other estimators, (3) and (4), provide unbiased results
(Figures~\ref{fig:PFsim} (d) and (e)). We also measured
$PF_{\rm rms}$ in equation~\ref{eq:PFrms} without
the small bias-correction term, $\sigma^2_{ak} + \sigma^2_{bk}$ which
is to remove the underlying Fourier power produced by noise
in the pulse profile. We find that this term changes the
results only by $\sim$0.1\% (see Table~\ref{ta:PFsim}).

We carried out simulations with different parameters
($A$ and $B$) for the theoretical pulse profile and find that
the results change depending on the number of events and
the pulse fraction. For example, the estimators (1) and
(2) tend to provide better results as the total count or
the pulsed fraction increases. However, the upward bias does
not disappear even for a fairly large number of total counts
(e.g., 10,000) or pulsed fraction (e.g., 70\%). The other
estimators did not significantly bias the results
in any set of parameters we studied.

We investigate effects of binning as the results can
change depending on binning. We binned the light curve into
8, 16, 32, 64, 100, and 128 bins. Note that we changed the total
number of events for different binning to keep the average
number of counts per bin same and to have similar statistical
error for the minimum (and maximum) phase bin. The results are shown in
Figure~\ref{fig:PFsim} (f). We find that estimators (1) and (2)
bias the results larger as the number of bins increases while
(3) and (4) produce robust results regardless of binning. This
is expected as there are more bins among which we can choose
the minimum when we increase the number of bins,
and hence it is more likely for estimators (1) and (2)
to have $p_{\rm min}$ smaller than the true minimum.
Estimators (3) and (4) do not rely on one minimum phase bin
but on the statistical average of the DC component,
and thus they are insensitive to the number of bins used. Furthermore,
they provide more accurate results as the number of bins increases
simply because we simulated more events in the cases of finer binning.

We find that estimator (3) provides the best results;
the measurements are closest to the theoretical value, and the
spread is smallest. Note that we had a priori knowledge on
the pulse shape with which we fit the pulse profile for
that estimator. Furthermore, the simple pulse shape allows
us to fit the profile only with two parameters. If the pulse
shape were more complex, one may have to use more harmonics
to fit the pulse profile and the results may be similar
to those obtained using $PF_{\rm rms}$
(e.g., see Figure~\ref{fig:pulsedFrac},
for error bars in two-harmonic fits). Hence, it may be easier to use
$PF_{\rm rms}$ estimator for more complex pulse profiles,
although the value measured with this estimator is different from
that of the others in general, and the conversion factor
(e.g., $\sqrt{2}$ for the sine function in the above simulation) can
change depending on the pulse shape, making direct comparison
with the others difficult.

In summary, we find that the pulsed fraction estimators
$PF_{\rm area}$ and $PF_{\rm minmax}$, often used in literatures, give
a biased result in general and are sensitive to the number of bins used.
Therefore, they should be used with great care.
The other estimators, $PF_{\rm fit}$ and $PF_{\rm rms}$, provide
an accurate measurement regardless of binning,
and hence they are preferred. We note, however,
that results of $PF_{\rm rms}$ cannot be directly
compared with those of the others since there is a scale factor
which differs for different pulse shape; the factor can be
obtained using simulations.

One final remark we would like to make is that extra care
needs to be taken when comparing pulsed fractions
measured with different instruments. In particular, if the pulse
fraction changes strongly over the energy band it was
measured, the result should be regarded as an energy-weighted
pulsed fraction. The energy-weighted pulsed fraction
will be measured differently by different instruments
since they have different energy responses. In this case, one has
to use either a smaller energy range over which the pulsed fraction
does not change much or a response-unfolded
estimator such as flux density ratio (A13).
\end{appendix}
\end{document}